\documentclass[twocolumn]{aastex62}

\usepackage{amsmath}
\usepackage{amssymb}
\usepackage{amsthm}
\usepackage{natbib,aasdefs,url,bm}
\usepackage{graphicx}
\usepackage{color}
\usepackage{mathtools,  epsfig, lipsum}
\usepackage{subfigure}
\usepackage{textcomp}

\def\beq{\begin{equation}}
\def\eeq{\end{equation}}
\def\bey{\begin{eqnarray}}
\def\eey{\end{eqnarray}}

\newcommand\T{\rule{0pt}{2.6ex}}       
\newcommand\B{\rule[-1.2ex]{0pt}{0pt}} 

\shorttitle{High-energy Cosmic Rays and Neutrinos from AT2018cow}
\shortauthors{Fang et al.}

\begin{document}

\title{Multimessenger Implications of AT2018cow: High-Energy Cosmic Ray and Neutrino Emissions from Magnetar-Powered Super-Luminous Transients}
\author{Ke Fang}
\affil{Einstein Fellow, Kavli Institute for Particle Astrophysics and Cosmology (KIPAC), Stanford University, Stanford, CA 94305, USA}

\author{Brian D.~Metzger} 
\affil{Department of Physics and Columbia Astrophysics Laboratory, Columbia University, New York, NY 10027, USA}

\author{Kohta Murase}
\affil{Department of Physics, The Pennsylvania State University, University Park, Pennsylvania 16802, USA}
\affil{Department of Astronomy \& Astrophysics, The Pennsylvania State University, University Park, Pennsylvania 16802, USA}
\affil{Center for Particle and Gravitational Astrophysics, The Pennsylvania State University, University Park, Pennsylvania 16802, USA}
\affil{Yukawa Institute for Theoretical Physics, Kyoto, Kyoto 606-8502 Japan}

\author{Imre Bartos}
\affil{Department of Physics, University of Florida, Gainesville, FL 32611, USA}
\affil{Department of Physics and Columbia Astrophysics Laboratory, Columbia University, New York, NY 10027, USA}

\author{Kumiko Kotera}
\affil{Sorbonne Universit\'es, UPMC Univ. Paris 6 et CNRS, UMR 7095,  Institut d'Astrophysique de Paris, 98 bis bd Arago, 75014 Paris, France}
\affil{Laboratoire AIM-Paris-Saclay, CEA/DSM/IRFU, CNRS, Universite Paris Diderot,  F-91191 Gif-sur-Yvette, France}

\begin{abstract}
Newly-born, rapidly-spinning magnetars have been invoked as the power sources of super-luminous transients, including the class of ``fast-blue optical transients" (FBOTs).  The extensive multi-wavelength analysis of AT2018cow, the first FBOT discovered in real time, is consistent with the magnetar scenario and offers an unprecedented opportunity to comprehend the nature of these sources and assess their broader implications. 
Using AT2018cow as a prototype, we investigate high-energy neutrino and cosmic ray production from FBOTs and the more general class of superluminous supernovae (SLSNe). By calculating the interaction of cosmic rays and the time-evolving radiation field and baryon background, we find that particles accelerated in the magnetar wind may escape the ejecta at ultrahigh energies (UHE). The predicted high-energy neutrino fluence from AT2018cow is below the sensitivity of the IceCube Observatory, and estimates of the cosmically-integrated neutrino flux from FBOTs are consistent with the extreme-high-energy upper limits posed by IceCube. 
High-energy $\gamma$ rays exceeding GeV energies are obscured for the first months to years by thermal photons in the magnetar nebula, but are potentially observable at later times.  Given also their potentially higher volumetric rate compared to other engine-powered transients (e.g.~SLSNe and gamma-ray bursts), we conclude that FBOTs are favorable targets for current and next-generation multi-messenger observatories.  
\end{abstract}

\section{Introduction}

Time-domain optical surveys have in recent years discovered a growing number of luminous, rare and rapidly-evolving extragalactic transients, with visual light curves that rise and decay on a timescale of only a few days (hereafter ``fast blue optical transients", or FBOTs; e.g.~\citealt{Drout+13,Drout+14,Arcavi+16,Rest+18}).  Although supernova light curves are often powered by the radioactive decay of $^{56}$Ni, such an energy source is ruled out for FBOTs (even if the entire ejecta mass were composed of nickel, the resulting optical luminosity would be insufficient to explain the observations). Instead, FBOTs must be powered by an additional, centrally-concentrated energy source or sources, either due to shock interaction between the ejecta and a massive circumstellar shell (e.g.~\citealt{Kleiser&Kasen18}), or from a central compact object, such as a rapidly spinning magnetar or accreting black hole.  

Indeed, newly-formed magnetars with millisecond rotation periods have widely been invoked as  engines responsible for powering the broader class of ``superluminous supernovae" (SLSNe; e.g.~\citealt{Kasen&Bildsten10,Woosley10}).  
Within this framework, the faster evolution of FBOTs requires stellar explosions with lower ejecta masses than of typical SLSNe.  Such low ejecta masses $\lesssim 0.1-1M_{\odot}$ could plausibly arise in a number of scenarios: explosions of ultra-stripped progenitor stars (e.g.~\citealt{Tauris+15}); electron capture supernovae (e.g.~\citealt{Moriya&Eldridge16}); initially ``failed" explosions of very massive stars such as blue supergiants (in which most of the stellar mass collapses into the central black hole, with only a tiny ejected fraction; e.g.~\citealt{Fernandez+18}); the accretion-induced collapse of a rotating white dwarf (e.g.~\citealt{Metzger+08, 2018arXiv181207569L}); magnetars created as the long-lived stable remnants of neutron star binary mergers (e.g.~\citealt{Yu+13,Metzger&Piro14,2018ApJ...854...60M}).  
On the other hand, compared to normal SLSNe or neutron star mergers, FBOTs may be relatively common, occurring at rates of up to $\sim 4-7$ \% that of the core collapse supernova rate (\citealt{Drout+14}). FBOTs also seem to occur exclusively in star-forming galaxies, supporting an origin associated with the deaths of massive stars \citep{Drout+14}. Rapidly-spinning pulsars in ultra-stripped supernovae have also been considered to explain some of the FBOTs~\citep{2017ApJ...850...18H}.

Until recently, all FBOTs were discovered at high redshift ($z > 0.1$) and were not identified in real-time, thereby precluding key spectroscopic or multi-wavelength follow-up observations.  This situation changed dramatically with the recent discovery of the optical transient AT2018cow     \citep{Smartt+18,CowATLAS}, which shared many characteristics with FBOTs but took place at a distance of only $\approx$ 60 Mpc ($z = 0.0141$).  AT2018cow reached a peak visual luminosity of $L_{\rm opt} \gtrsim 10^{44}$ erg s$^{-1}$ within only a few days, before fading approximately as a power-law $L_{\rm opt} \propto t^{-\alpha}$ with $\alpha \approx 2.5$ (e.g.~\citealt{Perley+18}).  The rapid rise time was indicative of an explosion with a low ejecta mass, $M_{\rm ej} \lesssim 0.5M_{\odot}$, and a high ejecta velocity, $v_{\rm ej} \gtrsim 0.1$ c.  The latter also revealed itself through the self-absorbed radio synchrotron emission from the source (\citealt{deUgartePostigo+18,Ho+18,CowX}), which was likely produced by the same fast ejecta undergoing shock interaction with a dense surrounding medium.  The early spectra were nearly featureless (also consistent with a high photospheric expansion speed), but after a few weeks narrow helium and hydrogen emission lines appeared \citep{Perley+18}, providing clues to the nature of the progenitor star (e.g.~\citealt{CowX}).  The persistent blue colors of the transient \citep{Perley+18}, again unlike the behavior of normal supernovae, provides additional evidence that a central ionizing energy source is responsible for indirectly powering the optical light through absorption and ``reprocessing" \citep{CowX}.  

Perhaps the most remarkable thing about AT2018cow is the direct visibility of the engine itself.  AT2018cow was accompanied by bright X-ray emission (\citealt{RiveraSandoval+18,CowX,Kuin+18}) with a power-law spectrum and luminosity comparable to that of the optical emission, but showing rapid variability on timescales at least as short as a few days.  The fact that X-rays appear to be visible from the same engine responsible for powering the optical luminosity provides evidence that the ejecta is globally asymmetric, such that the observed X-rays preferentially escape along low-density polar channels (see \citealt{CowX} for additional evidence and discussion).  \citet{CowATLAS} proposed that the engine behind AT2018cow is a millisecond magnetar with a dipole field strength of $B_{\rm d} \sim 10^{15}$ G and an initial spin period of $P_i\sim 10$ ms.  However, other magnetar parameters are consistent with the data and alternative central engine models remain viable, such as the collapse of a blue super-giant star to an accreting black hole, or a deeply embedded shock arising from circumstellar interaction \citep{CowX}.      

If AT2018cow is powered by a central compact object, particularly a millisecond magnetar, then it could also be a source of high energy charged particles \citep{Blasi00,Arons03,FKO12}, high-energy neutrinos \citep{2009PhRvD..79j3001M, 2014PhRvD..89d3012M,2014PhRvD..90j3005F, 2015JCAP...06..004F}, high-energy gamma-rays \citep{Kotera+13,2015ApJ...805...82M,Renault18}, and gravitational waves \citep{2005ApJ...634L.165S,2009MNRAS.398.1869D,2016ApJ...818...94K}.  Such a scenario has also been discussed in the context of pulsar-powered supernova, and the similar low-ejecta mass context of binary neutron star mergers (\citealt{Piro&Kollmeier16,Fang&Metzger17}). 
Indeed, two IceCube neutrino track events were reported in spatial coincidence with AT2018cow during a 3.5 day period following the optical discovery. The events are consistent with the rate of atmospheric neutrino background (\citealt{Blaufuss18}). While we will show that it is unlikely these neutrinos are physically associated with the transient, they nevertheless motivate a more thorough study of FBOTs as sources of ultrahigh energy cosmic rays (UHECRs) and neutrinos, particularly given their high volumetric rate as compared to other previously considered UHECR sites, such as gamma-ray bursts and SLSNe.
  
In this paper, we follow the procedure outlined in \citet{Fang&Metzger17} to calculate the neutrino emission from millisecond magnetars embedded in low ejecta mass explosions, focusing on parameters motivated by the watershed event, AT2018cow.  In \S\ref{sec:photonField} we describe our model for the radiation and hadronic background.  In \S\ref{sec:acceleration} we describe the acceleration and escape of UHECRs.  In \S\ref{sec:flux} we discuss neutrino production and address the neutrino detection prospects, both of AT2018cow and the cosmic background of FBOTs.  Throughout this paper we adopt the short-hand notation $q_{x} \equiv q/10^{x}$ cgs.

\section{Radiation and Hadron Background}
\label{sec:photonField}
The energetic compact object is embedded within a rapidly expanding ejecta shell filled with baryons and photons. Cosmic rays, which are accelerated close to the compact object, will lose energy and produce neutrinos when traveling through this dense medium. 
We follow \citet{Metzger&Piro14} to calculate the density of ejecta baryons, thermal and non-thermal photons.  This simplified model assumes the ejecta to be homogeneous and spherically symmetric.  While this is reasonable for estimating the environment for cosmic-ray interaction, modeling of AT2018cow indicates that the true ejecta structure is aspherical and thus is more complicated in detail (e.g.~\citealt{CowX, 2041-8205-868-2-L24}).

The evolution of non-thermal radiation $E_{\rm nth}$, thermal radiation $E_{\rm th}$, magnetic energy $E_B$, ejecta radius $R_{\rm ej}$ and velocity $v_{\rm ej}$ are described by a coupled set of differential equations:
\bey 
\frac{dE_{\rm nth}}{dt} &=& L_{\rm sd} -\frac{E_{\rm nth}}{R_{\rm n}}\frac{dR_{\rm n}}{dt} - \frac{E_{\rm nth}}{t_d^n}, \label{eqn:dE_nthdt} \\
\frac{dE_{\rm th}}{dt} &=& \frac{E_{\rm nth}}{t_d^n} -\frac{E_{\rm th}}{R_{\rm ej}}\frac{dR_{\rm ej}}{dt} -\frac{E_{\rm th}}{t_d^{\rm ej}}, \label{eqn:dE_thdt} \\
\frac{dE_B}{dt} &=& \epsilon_B\,L_{\rm sd} - \frac{E_B}{R_{\rm n}}\,\frac{dR_{\rm n}}{dt} \label{eqn:dE_Bdt}  \\
M_{\rm ej}v_{\rm ej}\frac{dv_{\rm ej}}{dt} &=& \frac{E_{\rm th} + E_{\rm nth} + E_B}{R_{\rm ej}}\frac{dR_{\rm ej}}{dt} , \label{eqn:dvdt} \\
\frac{dR_{\rm ej}}{dt} &=&  v_{\rm ej} \label{eqn:Rej}
\eey
The system is powered by the dipole spin-down of the magnetar \citep{1969ApJ...157.1395O}, which injects a luminosity given by
\bey
L_{\rm sd} & =& L_{\rm sd, 0} \,\left(1+\frac{t}{t_{\rm sd}}\right)^{-2}
\label{eqn:Lsd}
\eey
where $L_{\rm sd,0}=4 {\mu^2\Omega^4}/{9\,c^3} =2.6\times10^{45}\,B_{d,15}^2\,P_{i,-2}^{-4} \,\rm erg\,s^{-1}$ is the initial spin-down power\footnote{We use the vacuum dipole formula, following the precedent of \citet{2010ApJ...724L.199O}.  In reality the spin-down is instead best described as a force-free MHD wind (e.g.~\citealt{Spitkovsky06}), but the results are similar to within a factor of a few (e.g.~\citealt{2016ApJ...818...94K}).}, $\mu = B_d R_*^3$ is the magnetic moment of a pulsar with dipole field $B_d$ and stellar radius $R_*$, and $P_{i}=10^{-2}\,P_{i,-2}$~s is the initial spin period. The spin-down time, $t_{\rm sd}$, which characterizes the timescale over which most of the initial rotational energy, $E_{\rm rot} = I\Omega_i^2/2$, is released is given by
\beq
t_{\rm sd} \equiv \frac{E_{\rm rot}}{L_{\rm sd, 0}} = 1.0 \,P_{i,-2}^2\,B_{d, 15}^{-2}\,\rm d.
\eeq
The pulsar moment of inertia is given by $I=2\,M_* R_*^2\,/5$, where $R_{*} = 10$ km and $M_{*} = 1.4\,M_{\odot}$ are its radius and mass, respectively.  At times $t\gg t_{\rm sd}$, $L_{\rm sd}\propto t^{-2}$, corresponding to a pulsar braking index of $3$ \citep{1969ApJ...157.1395O}.  This decay rate is moderately shallower than that of the bolometric luminosity of AT2018cow ($L \propto t^{-2.5}$; \citealt{Perley+18,CowX}); however, a smaller braking index, or additional mechanisms such as the interaction of fall-back from the explosion with the magnetosphere \citep{0004-637X-857-2-95}, can result in a steeper decay. We also consider magnetic energy losses due to the expansion, which can be present, e.g., if the magnetic field in the nebula is turbulent.

Finally, internal magnetic fields at the scale of $10^{16}$~G can be generated in the core of newborn magnetars if their initial spin period is on the order of a few milliseconds \citep{Stella_2005, 10.1111/j.1365-2966.2008.14054.x}. The anisotropic pressure from the toroidal B-field leads to an ellipticity of $\varepsilon_{B, \rm GW} \sim 6.4\times10^{-4}\,B_{t,16.3}^2$ for an average toroidal field strength $B_t$ \citep{Stella_2005}. Such a quadrupole moment will cause a spin down due to gravitational wave emission at the rate $(d\Omega/ dt)_{\rm GW} = (32/5)(G/c^5)I\varepsilon^2_{B,\rm GW}\,\Omega^5$. In comparison, the spin-down rate due to the dipole field is $(d\Omega/dt)_{\rm d} = \mu^2 / (6Ic^3)\Omega^3$. The ratio of the two rates is $(d\Omega/ dt)_{\rm GW}/ (d\Omega/dt)_{\rm d} = 5\%\,P_{i,-3}^{-2}\,B_{d,15}^{-2}\,B_{t,16.3}^{2}$. This fraction could be higher depending on the equation of state of the stellar interior, but should not dominate the energy losses. We thus ignore the quadrupole term in this work, and note that it could impact the spin-down at the level of a few to ten percents. We note that an associated gravitational wave flux from such a source should not be detectable with current instruments \citep{Arons03, Kotera2011} except that it is extremely close \citep{2016ApJ...818...94K}, but a stochastic gravitational wave signature from a population of magnetars-powered transients could lead to detection with future instruments \citep{Kotera2011}.

A significant portion of the magnetar's rotational energy is ultimately used to accelerate the ejecta, as captured by equation~\ref{eqn:dvdt}. The mean velocity of the ejecta after time $t$ is thus approximately given by
\bey  
v_{\rm ej} &\approx&  \left(\frac{2\,\int_0^t\,L_{\rm sd}dt'}{M_{\rm ej}} + v_{\rm 0}^2\right)^{1/2},
\label{eq:betaest}
\eey
where $v_{\rm 0}\lesssim   0.05\,c$ is the initial ejecta velocity from the supernova explosion \citep{CowATLAS} and $M_{\rm ej} = 0.3\,M_{\rm ej,-0.5}\,M_\odot$ is the ejecta mass. 
The nebular radius $R_{\rm n}$ is taken to be a fixed fraction of the mean ejecta radius $R_{\rm n} = R_{\rm ej} / 5$, where $R_{\rm ej} \approx v_{\rm ej}t$, consistent with observations indicating that the photosphere velocity of AT2018cow declines from $\gtrsim 0.05\,c$ at $4$ days to $\sim 0.01-0.02\,c$ over its first two weeks of evolution \citep{CowATLAS}.

\begin{figure}
\includegraphics[width= \linewidth] {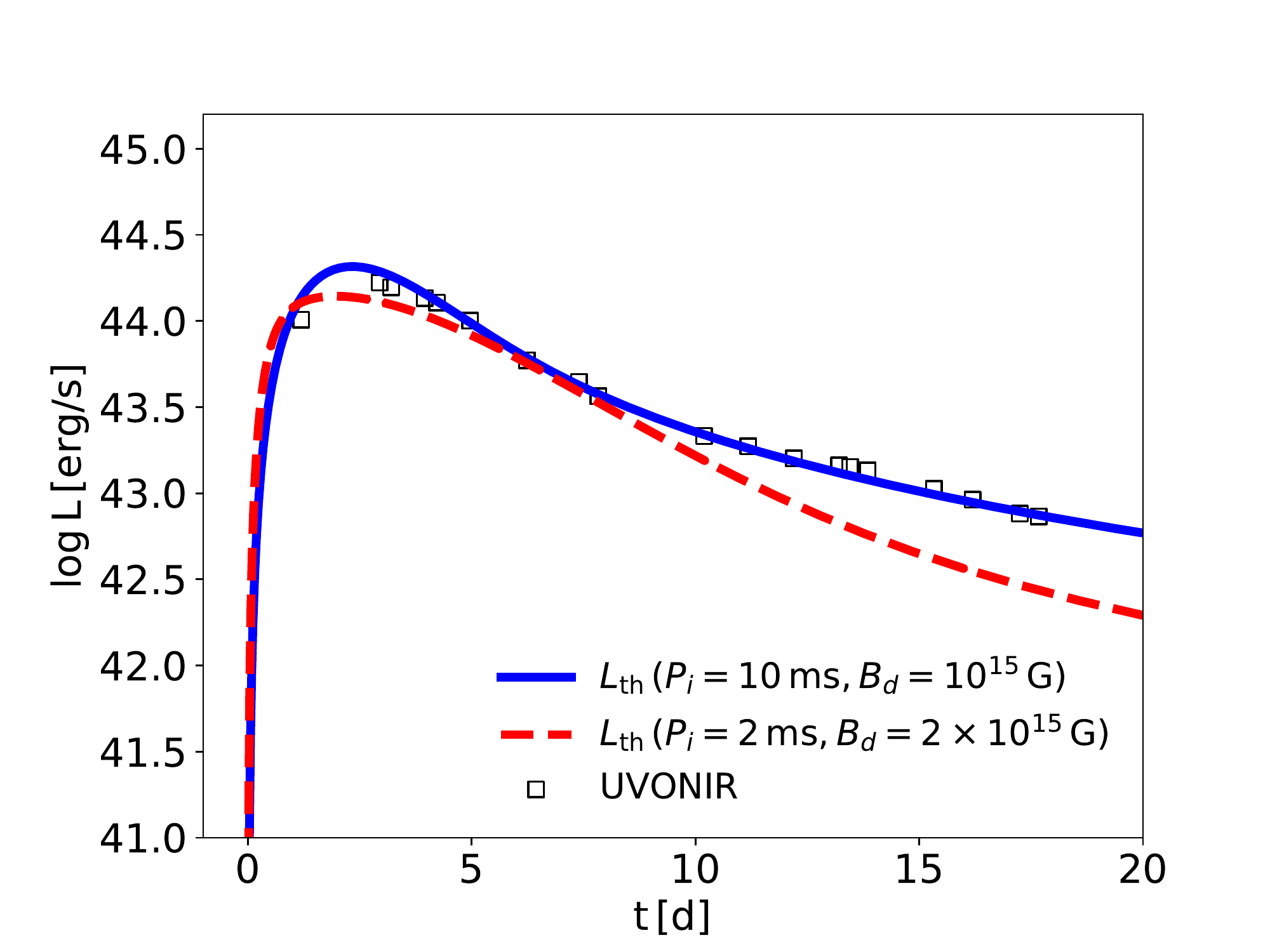}  
\caption{\label{fig:lightCurve}
Luminosity of thermal emission from the magnetar-powered transient (as calculated from eq.~\ref{eqn:L_th}) in comparison to the UVONIR light curve of AT2018cow     \citep{CowATLAS}.  Model results are shown separately for a magnetar with an initial rotation period $P_i=10\,\rm ms$, surface magnetic field $B_d=10^{15}\,\rm G$, ejecta opacity $\kappa = 0.01\,\rm cm^2\,g^{-1}$ (blue-solid line); and $P_i = 2\,\rm ms$, $B_d=2\times10^{15}$~G, $\kappa = 0.1\,\rm cm^2\,g^{-1}$ (orange-dashed line). In both cases, the ejecta mass is $M_{\rm ej} = 0.3\,M_\odot$ and the velocity of the magnetar-inflated nebula is taken to be five times lower than the mean ejecta velocity (to match the range of photosphere velocities observed in AT2018cow).}
\end{figure}

At the wind termination shock, the majority of the spin-down power is converted into non-thermal radiation~\citep[e.g.,][]{2010ApJ...715.1248T,2013MNRAS.429.2945T}. A portion of this is then converted to thermal radiation via absorption by the ejecta walls, as described by equations~\ref{eqn:dE_nthdt} and \ref{eqn:dE_thdt}, respectively.  The timescales $t_d^n$ and $t_d^{\rm ej}$ are those required for photons to diffuse radially through the nebula or ejecta shell, respectively. Specifically, 
\beq
t_d^n \approx \frac{R_{\rm n}}{c}\left(1+\tau_{\rm es}^n\right),
\label{eq:tdn}
\eeq
where $\tau_{\rm es}^n =  \left(4\,Y\,L_{\rm sd}\,\sigma_T/\pi\,m_e\,c^3\,R_{\rm n}\right)^{1/2} \approx  17\,B_{15}\beta_{\rm ej,-1.3}^{-1/2} \\ t_{5}^{-1/2}P_{i,-2}^{-2}$ is the  optical depth due to Thomson scattering by electron positron pairs.  The latter are generated by a pair cascade process in the nebula due to its high compactness \citep{Metzger&Piro14,Fang&Metzger17}, where $Y \approx 0.1$ is the pair multiplicity (fraction of the spin-down power converted into pair rest mass).  Once the photon diffusion time becomes less than the nebula expansion time $t_{\rm d}^n=R_{\rm n}/\beta_{\rm n} c$, photons are able to travel freely through the nebula; this occurs after a characteristic time
\beq
t_{d,0}^n \approx 0.03\,B_{15}^{2}P_{i,-2}^{-4} \beta_{\rm ej, -1.3}\,\rm d. 
\eeq

Similarly, photons diffuse through the ejecta on a timescale
\beq
t_d^{\rm ej} \approx \frac{R_{\rm ej}}{c}\left(1+\tau_{\rm es}^{\rm ej} \right),
\eeq
where $\tau_{\rm es}^{\rm ej} = {3M_{\rm ej}\kappa}/{4\pi R_{\rm ej}^2}$ is the optical depth and $\kappa\sim 0.01-0.4\,\rm cm^2\,g^{-1}$ is the ejecta opacity at UV/visual frequencies.  The free escape of photons takes place once $t_d^{\rm ej} < R_{\rm ej}/\beta_{\rm ej} c$, as occurs after the time

\beq
t_{d,0}^{\rm ej} = 2.1\, M_{\rm ej, -0.5}^{1/2}\left(\frac{\kappa}{0.01\,\rm cm^2\,g^{-1}}\right)^{1/2}\beta_{-1.3}^{-1/2}\,\rm d. 
\eeq
Equations~\ref{eqn:dE_nthdt} and \ref{eqn:dE_thdt} were derived under the approximation that all of the non-thermal radiation is reprocessed into thermal radiation following its absorption by the ejecta \citep{Metzger+14}.  This is justified by the observational fact that the X-ray luminosity of AT2018cow (that escaping from the engine) is less than the optical luminosity (that thermalized) over the first several weeks of evolution (e.g.~\citealt{CowX}).  

\begin{table}[t!]
\caption{Magnetar parameters of fiducial models} \label{table:models}
\centering
\begin{tabular}{ccc}
\hline\hline 
\T & $P_i$ [ms]   & $B_d$ [G] \B  \\
\hline
Case I \T  & 10 & $10^{15}$ \\
Case II \T&  2 & $2\times10^{15}$ \B \\
\hline
\end{tabular}
\end{table}

The thermal luminosity (supernova light curve) is given by the loss term in equation (\ref{eqn:dE_thdt}),
\beq\label{eqn:L_th}
L_{\rm th} = \frac{E_{\rm th}}{t_d^{\rm ej}}
\eeq
At early times  ($t\ll t_{d}^n$), the kinetic term ($PdV$ work) dominates the energy loss, such that $E_{\rm nth}\sim L_{\rm sd,0} t$ and $E_{\rm th}\sim  L_{\rm sd,0}t/\tau_{\rm es}^n \propto t^{3/2}$. The luminosity thus scales $L_{\rm th} \sim E_{\rm th}/t_{\rm d}^{\rm ej} \propto t^{5/2}$ at $t\ll t_{d,0}^{\rm ej}$. At late times  ($t\gg t_{d,0}^{\rm n}$, $t_{d,0}^{\rm ej}$), $E_{\rm nth}\sim L_{\rm sd,0} t_{\rm sd} \left(t/t_{\rm sd}\right)^{-1}$ and $E_{\rm th}\sim E_{\rm nth}t_{d}^{\rm ej} / t_d^n \propto t^{-1}$. As a result, $L_{\rm th}$ declines $\propto t^{-2}$, i.e. following the magnetar spin-down luminosity. 

Figure~\ref{fig:lightCurve} compares the time-dependent thermal luminosity calculated from equation~\ref{eqn:L_th} to the observed optical light curve of AT2018cow (as integrated from the UV to near IR wavelength bands). We consider two models for the properties of the central magnetar, referred to hereafter as Case I and Case II, respectively (see Table~\ref{table:models}). Case I corresponds to a magnetar with dipole field $B_d = 10^{15}$~G and initial spin period $P_i = 10$~ms. This model reproduces well the observed optical light curve, as shown in Figure~\ref{fig:lightCurve}, and is consistent with previous fits to the magnetar model \citep{CowATLAS,CowX}.  We use parameters from Case I in our analytic estimates below.  Case II corresponds to a magnetar with a somewhat stronger field $B_d = 2\times 10^{15}$~G and larger rotational energy $P_i = 2$~ms. The light curve in this case, as shown by a dashed line in Figure~\ref{fig:lightCurve}, is also roughly consistent with that of AT2018cow (though decaying a bit too quickly at late times).

The magnetic field in the magnetar-wind nebula can be estimated as
\beq\label{eqn:B}
B_{\rm n} \approx \left(\frac{6\epsilon_B L_{\rm sd}t}{R_{\rm n}^3}\right)^{1/2}  \simeq 720\,\epsilon_{B,-4}^{1/2}\,B_{15}\,P_{i,-2}^{-2}\,\beta_{n,-2}^{-3/2}\,t_{5}^{-2}\,\rm G,
\eeq
where a fraction $\epsilon_{\rm B}$ of the spin-down energy $\sim L_{\rm sd}t$ is assumed to be placed into the magnetic energy of the nebula, $E_B = (B_n^2/8\,\pi)V_{\rm n}$ (e.g. at the wind termination shock), where $V_{\rm n} = 4\pi R_{\rm n}^{3}/3$ is the nebula volume.  Typical values $\epsilon_B\sim 10^{-4} - 10^{-2}$ are obtained by modeling pulsar wind nebulae (e.g, \citealt{1984ApJ...283..694K, TORRES201431}).  

The temperature of the ejecta, and thus that of the thermal radiation field, is approximately given by
$T_{\rm th} = \left({3 E_{\rm th}}/{4\pi a R_{\rm ej}^3}\right)^{1/4}$.
The number density of thermal photons is then given by
\begin{eqnarray}
n_{\rm th} &\simeq& 16\pi\zeta(3) \left({kT_{\rm th}}/{hc} \right)^3 \nonumber \\
&\approx& 2.3\times10^{16}\,\,B_{15}^{3/2} P_{i,-2}^{-3} t_{5}^{-3/2} \beta_{\rm ej,-1.3}^{-3/2}\,\rm cm^{-3} \ .
\end{eqnarray}

The thermal emission of the hot magnetar as considered in \cite{Kotera_2015} is subdominant here as the acceleration and interaction sites are distant from the star. As detailed in Section~\ref{sec:acceleration}, the acceleration should happen beyond the light cylinder. The density of thermal photons from the magnetar is a factor of $(R_{\rm star} / R_{\rm acc})^2$ times lower than that near the star, where $R_{\rm star}$ is the stellar radius and $R_{\rm acc}$ the radius of the acceleration site. The contribution of these photons to neutrino production is thus negligible. 

 The number density of non-thermal photons can likewise be estimated to be
\bey\label{eqn:nnth}
n_{\rm nth} &\sim& \frac{L_{\rm nth}}{4\pi R_{\rm n}^2 c\,\varepsilon_{\rm min}\,\ln\left(\varepsilon_{\rm max}/\varepsilon_{\rm min}\right)} \\ \nonumber
&\approx&3.6\times10^{15}\,B_{15}^{3/2}\,P_{i,-2}^{-3}\,t_{5}^{-3/2}\,\beta_{n,-2}^{-3/2}\,\rm cm^{-3},
\eey
where we have assumed a flat power-law spectrum $n_{\rm nth}(\varepsilon) \propto \varepsilon^{-2}$, extending from the energy of the thermal radiation $\varepsilon_{\rm min}\sim 3\,k_B\,T_{\rm th}$ to the pair creation threshold $\varepsilon_{\rm max}\sim 2\,m_e c^2$ \citep{1987MNRAS.227..403S}.  The observed X-ray spectrum of AT2018cow ($n(\varepsilon) \sim \varepsilon^{-1.5}$; \citealt{CowX}) is somewhat harder than assumed in the model.  However, this difference does not critically affect our conclusions because the density of the non-thermal radiation is lower than that of the thermal photons (as well as higher in energy), and therefore are generally less important targets for neutrino production.
 
The baryon density of the ejecta is given by
\beq\label{eqn:np}
n_p \approx \frac{3 M_{\rm ej}}{4\pi R_{\rm ej}^3\,m_p} \approx 2.5\times10^{13}\,M_{\rm ej,-0.5}\,t_{5}^{-3}\,\beta_{\rm ej,-1.3}^{-3}\,\rm cm^{-3}.
\eeq
While $n_p$ scales with time as $\propto t^{-3}$, the evolution of $n_{\rm th}$ changes at $t_{d,0}^{\rm n}$ and $t_{d,0}^{\rm ej}$, introducing features to the cosmic-ray interaction, as discussed below.

\section{Acceleration and Escape of UHECRs}
\label{sec:acceleration}
\subsection{Cosmic Ray Injection}
Pulsars and magnetars offer promising sites for particle acceleration. In general, charged particles may tap the open field voltage and gain an energy \citep{Arons03}   
\bey\label{eqn:ECR}
&&E_{\rm CR} \simeq \eta\,Z\,e\,\Phi_{\rm mag} \nonumber \\
&\approx& 1.3\times10^{19}\,Z\,\eta_{-1}\,B_{15}\,P_{i,-2}^{-2}\left(1+\frac{t}{t_{\rm sd}}\right)^{-1}\rm eV,
\eey
where $\eta=0.1\,\eta_{-1}$ is the acceleration efficiency and $Z$ is the particle charge. For simplicity, we assume cosmic rays of proton composition. The effects of a heavier composition on neutrino production are discussed in Section~\ref{sec:discussion}. 

Assuming that ions follow the Goldreich-Julian charge density, the rate of cosmic ray injection from the magnetic polar cap is given by $\dot{N} = \mu \Omega^2 / Ze c$ \citep{Arons03}. Equation~\ref{eqn:ECR} then implies the energy spectrum of accelerated particles is given by
\beq\label{eqn:dNdE}
\frac{dN}{dE} = \frac{9}{8}\frac{c^2\,I}{e\,\mu}\,\frac{1}{E} = 2\times10^{42}\,\,E^{-1}\,B_{15}^{-1}.
\eeq

The specific acceleration mechanism of high-energy particles in the pulsar magnetosphere is still debated (e.g.~\citealt{2017SSRv..207..111C}). For UHECRs, \citet{Arons03} hypothesized that particles achieve their energy in the relativistic wind through surf-riding acceleration at a radius $10^3-10^4$ times the light cylinder. 
The possibility of wake-field acceleration has also been discussed~\citep[e.g.,][]{2009PhRvD..79j3001M,2017ApJ...840...52I}. 
Recent particle-in-cell simulations \citep{2018ApJ...855...94P} support a significant fraction of the accelerated particle energy flux being carried by ions extracted from the stellar surface. The ion energy may reach up to $10-30\%$ of the polar cap voltage. As the misalignment of the rotational and magnetic axises impact the current distribution, the spectrum also depends on the inclination angle of the pulsar. Alternatively, if the acceleration happens close to the star, curvature radiation will be non-negligible and limit the maximum energy. However, particles could be picked up and re-accelerated by the pulsar wind. If the pair multiplicity is low, a significant wind power would go into ions and still allow a UHECR production \citep{Kotera_2015}.

\subsection{Interaction and Cooling Rates}\label{sec:timescales}
\begin{figure}[t!]
\includegraphics[width=  \linewidth] {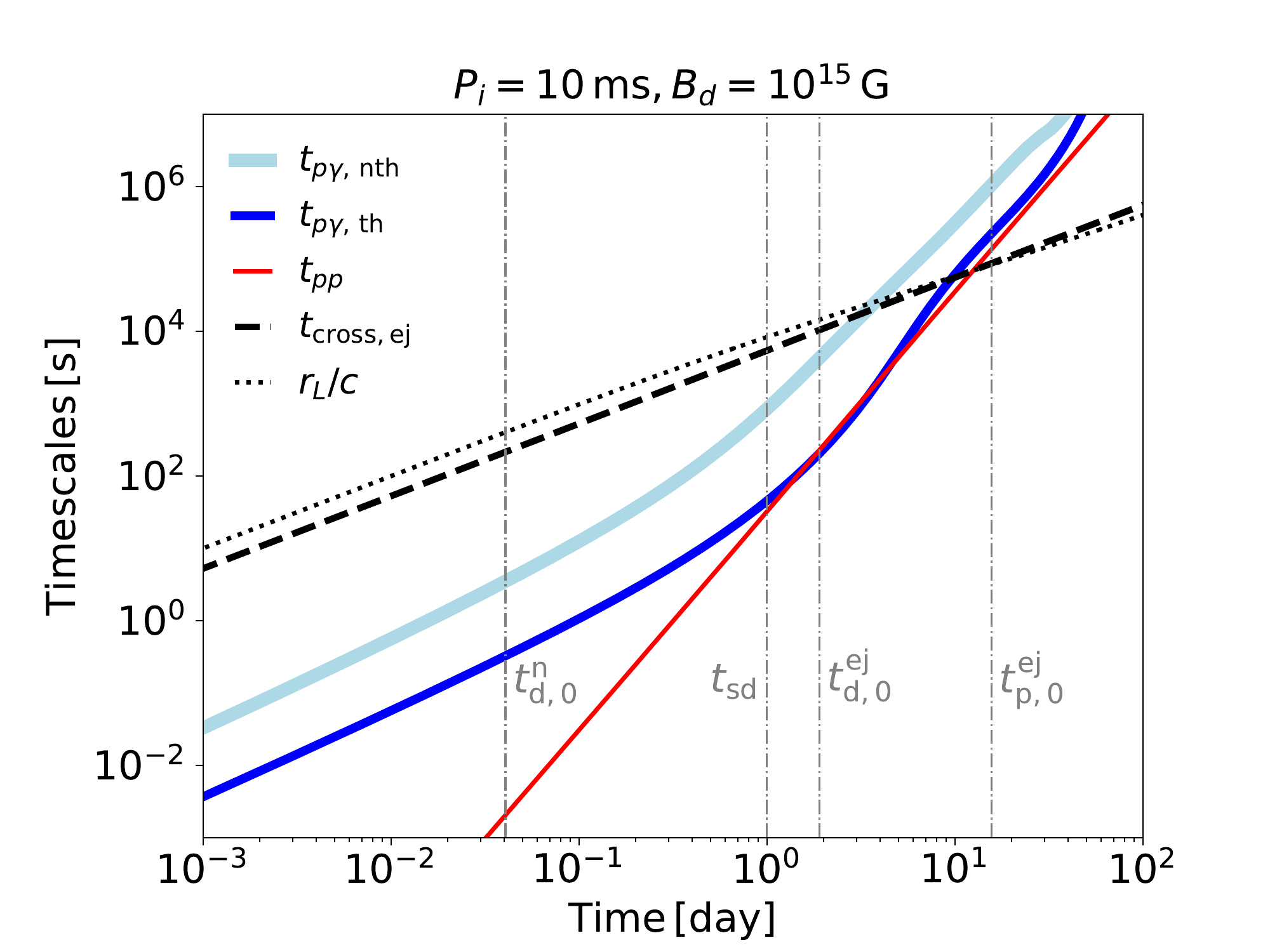}  
\includegraphics[width=  \linewidth] {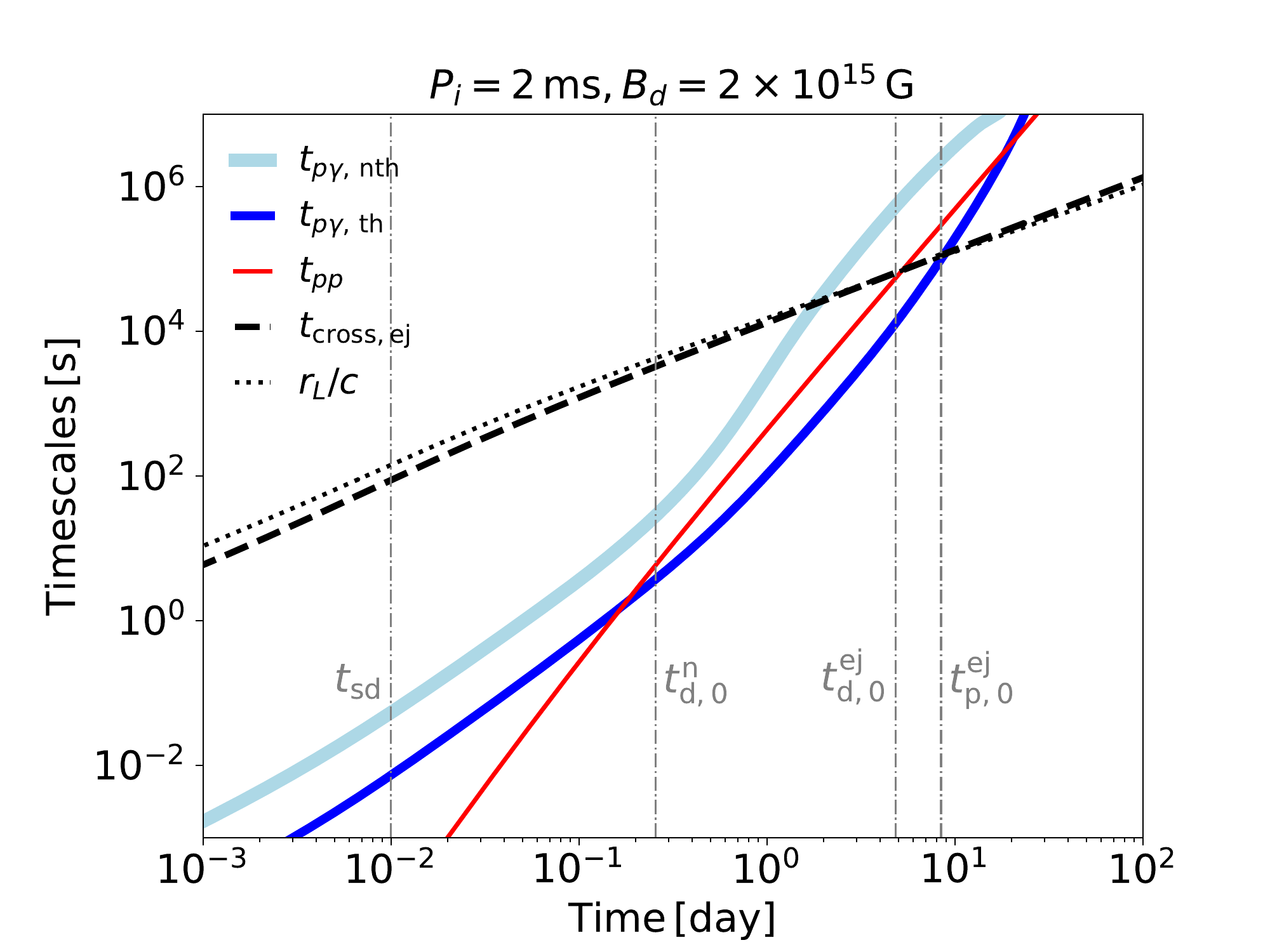}  
\caption{\label{fig:t_parents}
Rates of interaction and cooling processes of protons as a function of time since explosion.  The photopion production rates are computed using equation~\ref{eqn:t_pgamma} for the thermal (blue-solid) and non-thermal background (light blue-solid) density computed in Section~\ref{sec:photonField}. The hadron interaction rates are from equations~\ref{eqn:t_pp} (red-solid). The proton energy is determined by equation~\ref{eqn:ECR}. At all times, the particle's Lamor radius (black-dashed) is comparable to the size of the nebula (black-dotted), meaning that both synchrotron and adiabatic losses of the protons are negligible. The magnetar parameters are $P_i = 10$~ms and $B_d=10^{15}$~G in Case I (top) and $P_i = 2$~ms and $B_d=2\times10^{15}$~G in Case II (bottom). }
\end{figure}

Accelerated cosmic rays interact with the background of baryons or photons, producing charged pions ($p+p(\gamma)\rightarrow \pi^\pm$) which decay into muons and neutrinos ($\pi^\pm \rightarrow \mu^\pm + \nu_\mu(\bar{\nu}_\mu)$).  The muons further decay into electrons and neutrinos ($\mu^\pm \rightarrow e^\pm + \nu_\mu + \bar{\nu}_\mu + \nu_e (\bar{\nu}_e)$).  Each step of this process may be suppressed by particle cooling.

A proton with Lorentz factor $\gamma_p$ interacts with the photon field of spectrum $n(\varepsilon)=dn/d\varepsilon$ on a characteristic timescale given by
\bey\label{eqn:t_pgamma}
t_{p\gamma,\,\rm int}^{-1} = \frac{c}{2\gamma_p^2}\,\int_0^\infty\,d\varepsilon\,\frac{n(\varepsilon)}{\varepsilon^2}\,\int_0^{2\,\gamma_p\,\varepsilon}\,d\varepsilon'\,\varepsilon'\,\sigma_{p\gamma}(\varepsilon')
\eey
where $\sigma_{p\gamma} $ is the cross section of  photopion production. 
The cooling time   is  $t_{\,p\gamma} = t_{p\gamma,\,\rm int}/ \kappa_{p\gamma}\sim \left(n_\gamma\,\sigma_{p\gamma}\,\kappa_{p\gamma}\,c\right)^{-1}$, where $\sigma_{p\gamma}\kappa_{p\gamma}\sim 10^{-28}\,\rm cm^{-2}$  is the  inelastic component of the $p\gamma$ interaction cross section.  The rate for hadronuclear interaction is likewise given by
\beq \label{eqn:t_pp}
t_{pp}^{-1} = n_p\,\sigma_{pp}\,\kappa_{pp}\,c
\eeq
where $\sigma_{\rm pp}\sim 10^{-25}\,\rm cm^2$ (at energies $\sim 10^{18}$~eV) and $\kappa_{\rm pp}\sim 0.5$~\citep[e.g.,][]{2004PhLB..592....1E}. 
The total interaction rate of protons, due to both $pp$ and $p\gamma$ processes, is then $t_p ^{-1} = t_{pp}^{-1} + t_{p\gamma}^{-1}$. 

The gyro radii of protons in the magnetic field of the nebula, $r_L$, is comparable or larger than the nebula size, 
\beq 
\frac{r_L}{R_{\rm n}} \approx 1.7\,\eta_{-1}\,\epsilon_{B,-4}^{-1/2}\,\beta_{n,-2}^{1/2}, \quad t\gg t_{\rm sd}.
\eeq 
Synchrotron or adiabatic losses of cosmic rays crossing the nebula are thus generally only marginally important. Inverse Compton cooling at such high energies is furthermore suppressed and negligible due to the Klein-Nishina effect. 

Protons travel freely when their crossing time is shorter than the interaction time, $t_{\rm cross, ej} < t_p$, as occurs after a characteristic timescale
\beq
t_{p,0}^{\rm ej}\approx 16\,M_{\rm ej,-0.5}^{1/2}\,\beta_{\rm ej, -1.3}^{-1}\,,
\rm d.
\eeq
where we have assumed that at late times $pp$ interactions dominate over $p\gamma$ interactions (however, note that $p\gamma$ interactions are included in our full numerical calculations). 

Figure~\ref{fig:t_parents} shows the proton interaction timescales as a function of time since explosion. In Case I, the dominant process for protons is hadronuclear interaction. In Case II, the photopion production with the thermal photon background becomes important at late times, $t \gtrsim t_{d,0}^n$. In both cases, the ejecta becomes optically thin at roughly one week, after which time the accelerated protons freely escape to infinity.

\subsection{Ultrahigh Energy Cosmic Ray Production}
\begin{figure}[t!] 
\includegraphics[width=  \linewidth] {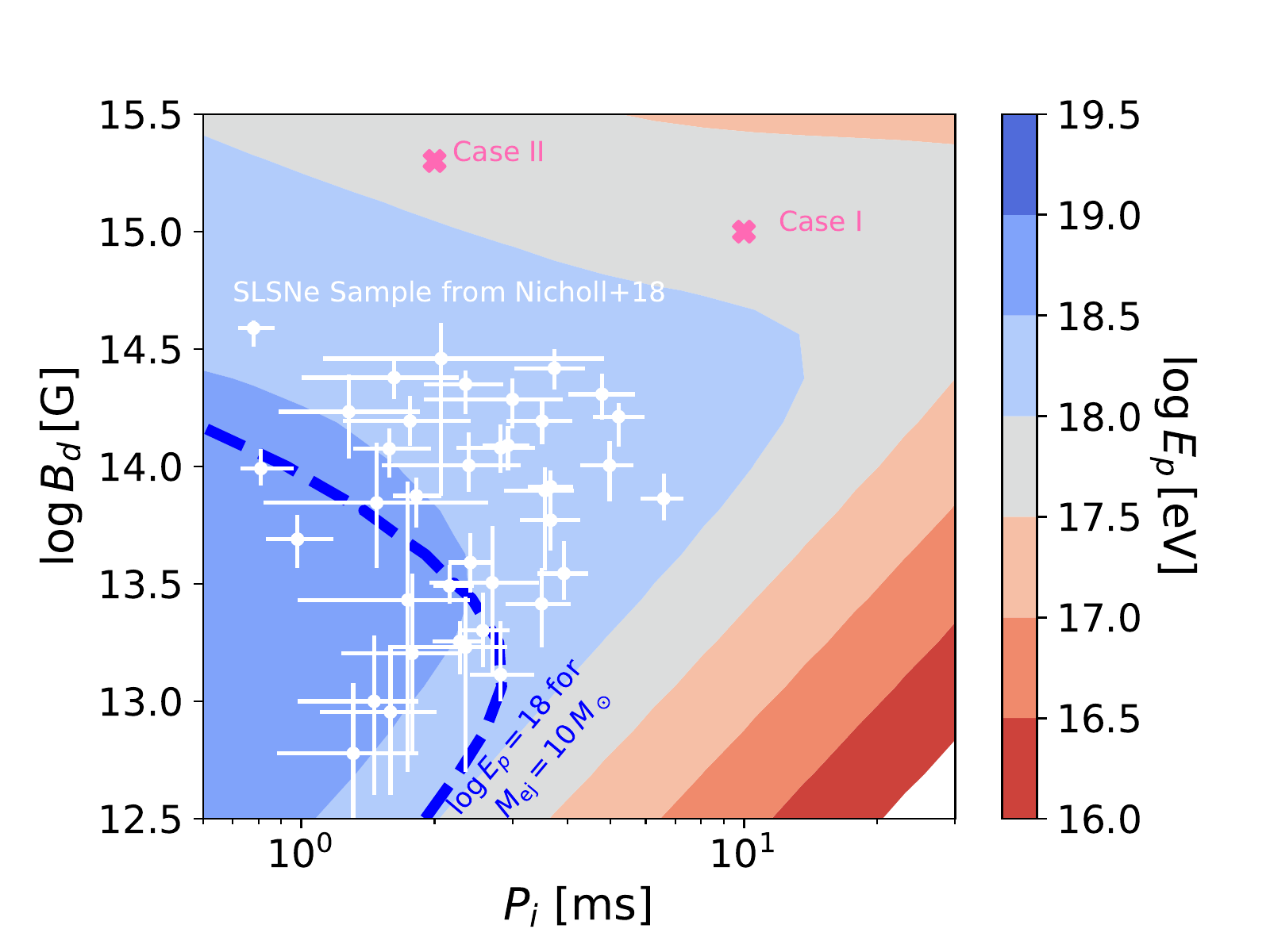}  
\caption{\label{fig:E_esc}
Maximum energy of cosmic rays at the time when the ejecta first become optically thin to $pp$ and $p\gamma$ interactions as a function of the surface magnetic field $B_d$ and birth period $P_i$ of the neutron star.  Magnetars with $P_i < 10$~ms are promising sources of UHE protons in transients with low ejecta masses $M_{\rm ej} \lesssim 0.3\,M_\odot$ such as AT2018-cow (the parameters of our fiducial models, Case I and II, are denoted as pink crosses).  By contrast, millisecond magnetars in standard SLSNe with larger ejecta mass $M_{\rm ej}=10\,M_\odot$ require values of $(B_d\,,P_i)$ to the left of the blue dashed curve to allow the escape of protons with $E_p \geq 10^{18}$~eV.  White points show magnetar parameters fit to a sample of SLSNe light curves by \citet{2017ApJ...850...55N}.  All calculations assume values of $\eta = 0.1$ and $\kappa = 0.1\,\rm cm^2\,g^{-1}$.}
\end{figure}

Figure~\ref{fig:E_esc} shows the maximum energy of the accelerated UHECR protons at the time when they escape the ejecta freely. The shaded area corresponds to an event with a low ejecta mass $M_{\rm ej} = 0.3\,M_\odot$, similar to that inferred for FBOTs such as AT2018cow.  Depending on the magnetic field of the magnetar, an AT2018cow-like event with $P_i \lesssim 10$~ms is a promising source for cosmic rays of energy $\gtrsim 10^{18}$~eV (see also \citealt{Piro&Kollmeier16}).  If the magnetar wind is composed of nuclei with charge $Z$ instead of protons ($Z = 1$), then their maximum energy is a factor of $Z$ times higher than for protons and thus in the range of the highest energy cosmic rays observed by Auger \citep{2017arXiv170806592T} and TA \citep{Matthews:2017waf}, as detailed in \citet{FKO12}.

For comparison, millisecond magnetars born in the normally considered class of superluminous supernovae (SLSNe), for which the ejecta mass is typically much higher ($M_{\rm ej} \sim 10\,M_\odot$), only allow the escape of UHE protons for magnetar parameters to the left of the blue-dashed curve.  This is because massive ejecta shells require longer to become optically thin, delaying the time of cosmic ray escape and thus reducing the pulsar voltage at this epoch.  White data points show magnetar parameters fit to the sample of SLSNe in \citet{2017ApJ...850...55N}, roughly one third of which appear to be promising UHECR sources. 

We conclude this section by estimating the UHECR energy budget of FBOTs similar to AT2018cow. Assuming $P_i = 2-10\,\rm ms$ and that a fraction $\eta = 0.1$ of the magnetar's rotational energy goes into cosmic rays, the UHECR energy yield per event is $\left(0.2-5\right)\times{10}^{50}$~erg.  The total rate of FBOTs is $\sim 4-7\%$ of the core-collapse supernovae rate \citep{Drout+14}.  However, the rate of the most luminous members of this class like AT2018cow is likely lower (M.~Drout, private communication), perhaps only $\sim 0.5-1\%$ of the core collapse supernova (CCSN) rate, corresponding to a estimated volumetric rate of AT2018cow-like events of $600-1200$\,Gpc$^{-3}\,$yr$^{-1}$.  
Using a FBOT rate with $\sim1$\% of the CCSN rate, the integrated cosmic ray luminosity density from AT2018cow-like FBOTs is thus roughly estimated to be $\sim\left(0.3-6\right)\times10^{44}\,\rm erg\,Mpc^{-3}\,yr^{-1}$. 
This is comparable to the UHECR luminosity density, which is in the order of $10^{44}\,\rm erg\,Mpc^{-3}\,yr^{-1}$~\citep[e.g.,][]{2018arXiv180604194M}.

\section{High-Energy Neutrino Production} \label{sec:flux}
\subsection{IceCube Observation of AT2018cow}
Two IceCube neutrino track events with $\sim 2^\circ$ angular resolution were found in spatial coincidence with AT2018cow during a $3.5$ day period between the last non-detection and the discovery, corresponding to a $1.8\,\sigma$ chance coincidence.  Assuming an $E^{-2}$ spectrum,  a time-integrated $\nu_\mu$ flux upper limit $\left(E_\nu^2\,J_\nu\right)_{\rm UL}^{\rm IC} = 6.1 \times10^{-2} \,\rm GeV\, cm^{-2}$ is found for this observation period at 90\% CL\footnote{http://www.astronomerstelegram.org/?read=11785}. 

Neutrinos with TeV-PeV energies lie in the best sensitivity window for the IceCube Observatory, and are predicted to arrive in the first few hours in our model. Using the neutrino effective area of the IceCube Observatory with its complete configuration of 86 string detectors \footnote{https://icecube.wisc.edu/science/data/PS-IC86-2011} \citep{2014ApJ...796..109A}, we compute the detector sensitivity at the declination of AT2018cow as a function of energy. It is shown by the grey dashed curve in Figure~\ref{fig:phi_neu}. Point sources with fluxes above the curve, assuming that they follow an $E^{-2}$ spectrum over a half decade in energy, are excluded at 90\% C. L.

\subsection{Competitive Cooling and Decay of Pions and Muons}

\begin{figure}[t!]
\includegraphics[width=  \linewidth] {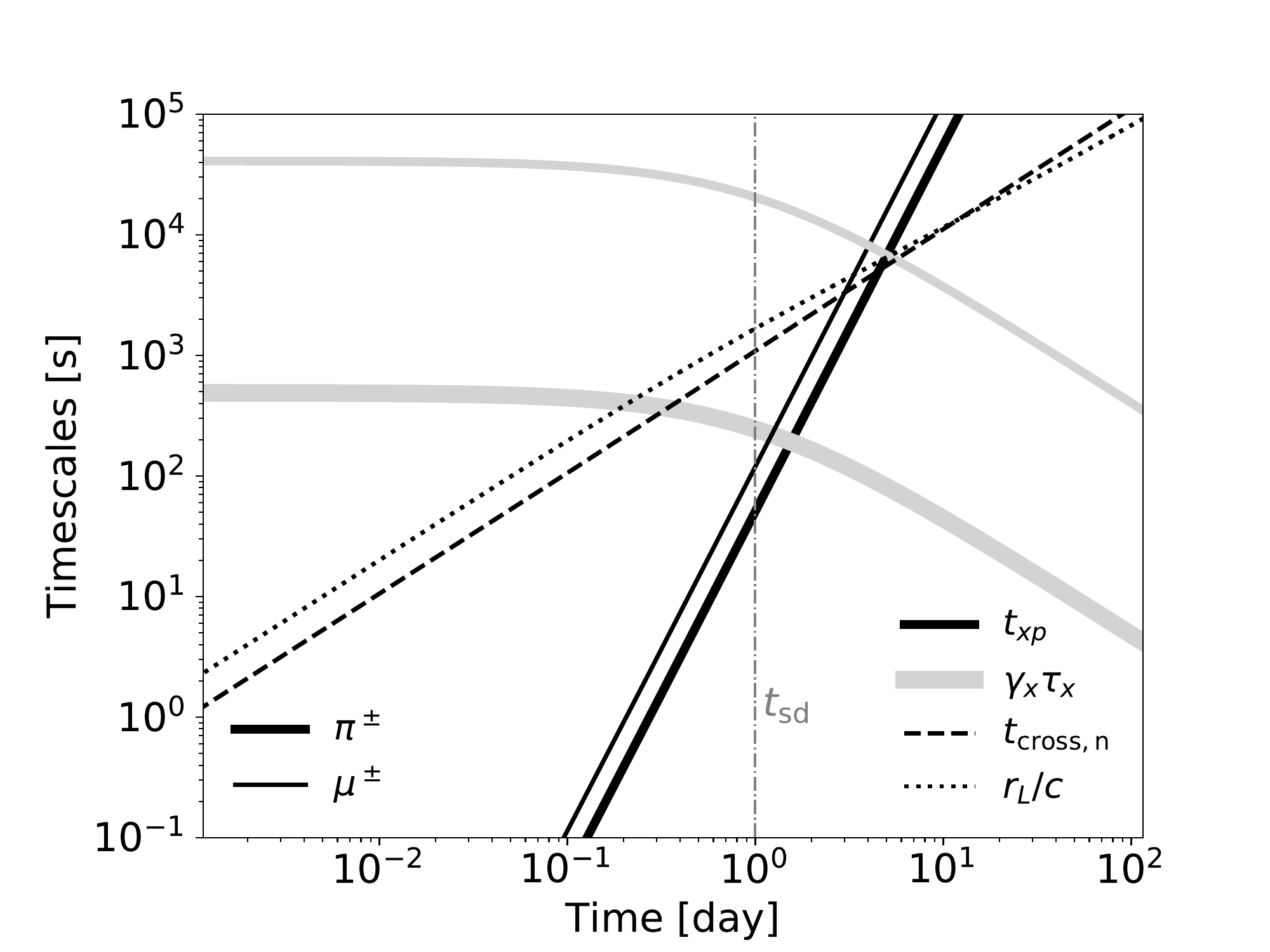}  
\includegraphics[width=  \linewidth] {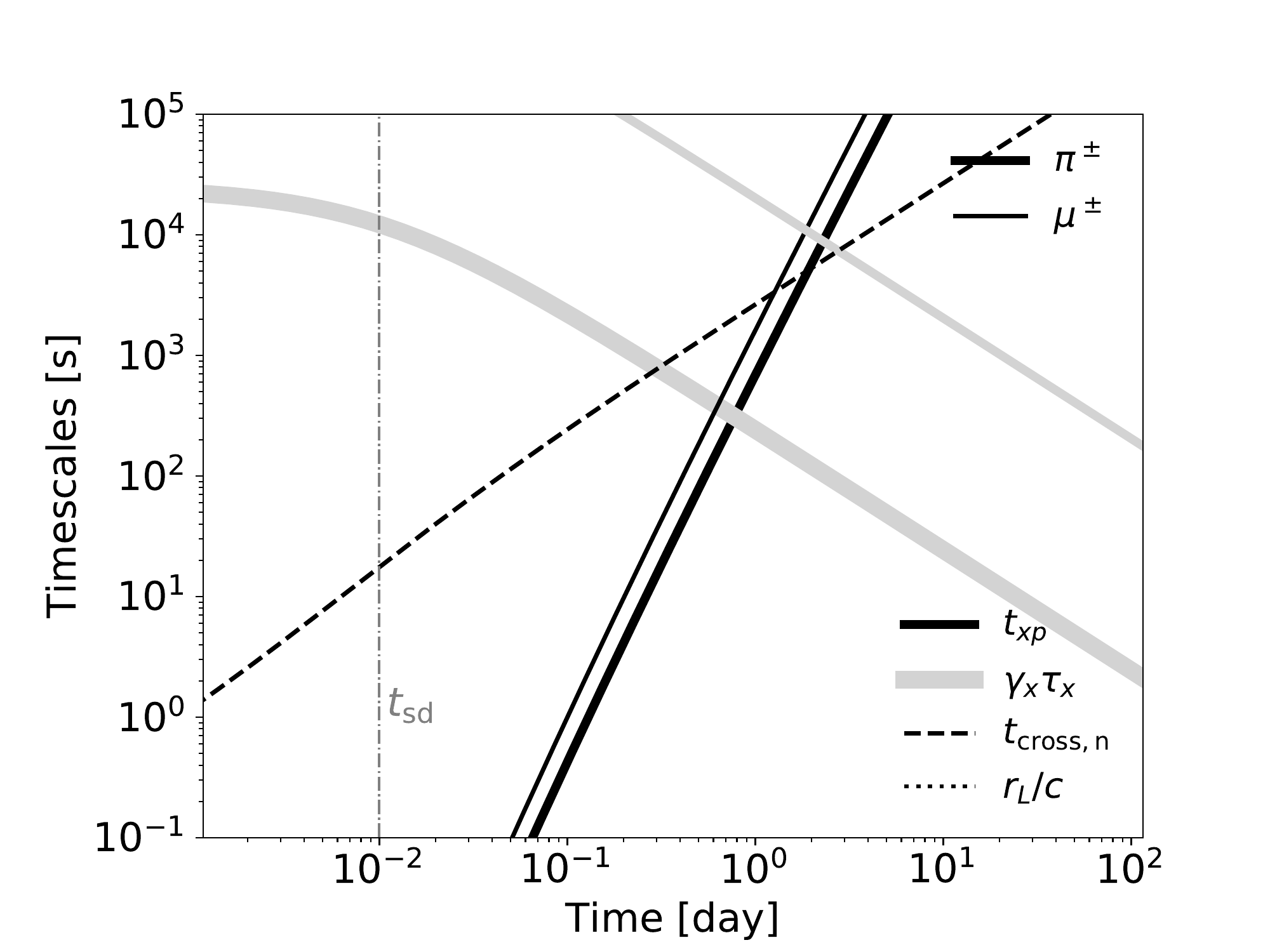}  
\caption{\label{fig:t_pion}
Interaction and decay rate of pions and muons as a function of time since explosion.  The pion and muon are assumed to carry $20\%$ and $15\%$, respectively, of the proton energy determined by equation~\ref{eqn:ECR}.  Pions undergo severe $\pi p$ losses at early times, but are free to decay after a day or so.  Due to their longer rest-frame lifetimes, muons experience strong cooling for a longer duration than pions. Like protons, the gyroradii of pions and muons in the magnetic field of the nebula are comparable to the nebula size and thus these particles are not subject to strong synchrotron losses unless the nebula magnetization is high, $\epsilon_B \gtrsim 10^{-2}$.}
\end{figure}

Charged pions and muons decay into neutrinos following their lab-frame lifetimes, $\gamma_x \,\tau_x$, where $x$ denotes either $\pi$ or $\mu$, and $\tau_{\pi^\pm} =2.6\times10^{-8}\,\rm s$ and $\tau_\mu =2.2\times10^{-6}\,\rm s$ are the rest-frame lifetimes.  These lifetimes can be long enough to allow cooling processes to occur.  Mesons and muons interact with the hadron background at a rate given by
\beq\label{eqn:t_xp}
t_{x p}^{-1} = n_p\,\sigma_{x p}\,\kappa_{x p}\,c.
\eeq 
Mesons and muons can also interact with the photon background, but as in the case of protons the smaller $\pi\gamma$ cross section renders the $\pi p$ interaction more important, especially at early times.   

Figure~\ref{fig:t_pion} compares the lifetime of a pion (muon), which is assumed to carry away 20\% (15\%) of the energy of the parent proton, to the cooling time due to $\pi p$ interactions.  
This shows that most pions interact with the ejecta baryons, rather than decay, in the first day or longer. 

Pion-proton interactions produce lower-energy pions and protons, which then undergo further interactions with background particles.  Eventually, higher-order pions reach sufficiently low energies to enable their decay into neutrinos.  This happens once $t_{\pi p} > \gamma_\pi \tau_\pi$, as occurs after a time
\beq 
t_{\pi p,0} \approx 2.2 \, \eta_{-1}^{1/4}\,B_{15}^{-1/4}\,\beta_{\rm ej, -1.3}^{-3/4}\,M_{\rm ej,-0.5}^{1/4}\,\rm d.
\eeq 
This effect introduces a break at time $t$ in the neutrino spectrum at a characteristic energy \citep{2009PhRvD..79j3001M}
\beq 
E_{\nu, b} \approx 59.7\,M_{\rm ej,-0.5}^{-1} \,t_5^{3}\,\beta_{\rm ej,-1.3}^3\,\rm PeV. 
\label{eq:Enub}
\eeq
In addition to the higher-order products, multiple pions and other types of mesons such as kaons may be produced from each $pp$ and $p\gamma$ interaction.  We take these into account in our numerical simulations presented in  Section~\ref{sec:numericalSetup}.  

Finally, similar to protons, the gyroradii of pions and muons are comparable to the nebula size.  Synchrotron cooling thus becomes important if $r_L / R_{\rm n} \ll 1$, as occurs for high nebular magnetization $\epsilon_B\gtrsim 10^{-2}$. 

\subsection{Numerical Procedure}\label{sec:numericalSetup}

We compute the neutrino flux from magnetar-powered supernovae with low-ejecta masses according to the following numerical procedure.  At each time $t$ following the explosion, the ejecta radius $R_{\rm ej}$, temperature of the thermal background $T_{\rm th}$, number density of ejecta baryons $n_b$, and strength of the nebular magnetic field $B_n$, are calculated from equations~\ref{eqn:dE_nthdt}$-$\ref{eqn:Rej}. Protons of energy $E_p$ determined by equation~\ref{eqn:ECR} are injected into the ejecta.  Their interaction with the baryon ejecta is calculated using a Monte-Carlo approach employing the hadronic interaction model EPOS-LHC~\citep{PhysRevC.92.034906} as in \citet{FKO12}. 
Photomeson interactions between cosmic-ray protons and the thermal background is computed based on SOPHIA~\citep{2000CoPhC.124..290M} through CRPropa 3 \citep{2016JCAP...05..038A}. 

As we are particularly interested in the earliest phases of the transient (near the time of the observed neutrino coincidence), we take a different approach in treating the pion-proton ($\pi p$) interaction from previous works (e.g. \citealt{2009PhRvD..79j3001M, 2016JCAP...04..010F}). The suppression of neutrino production due to $\pi p$ interaction is usually described by a suppression factor $f_{\rm sup}\equiv \min\left(t_{\rm cool}/\gamma_\pi\,\tau_\pi, 1\right)$, with $\gamma_\pi$ and $\tau_\pi$ being the Lorentz factor and the rest-frame lifetime of the charged pion. This analytical approach however misses the secondary and higher-order pions produced by the interaction, which have lower energy and thus decay into neutrinos more easily than their parents.  This is a secondary effect, but can play an important role in neutrino production in dense environments.

\begin{figure}[t!]
\includegraphics[width=  \linewidth] {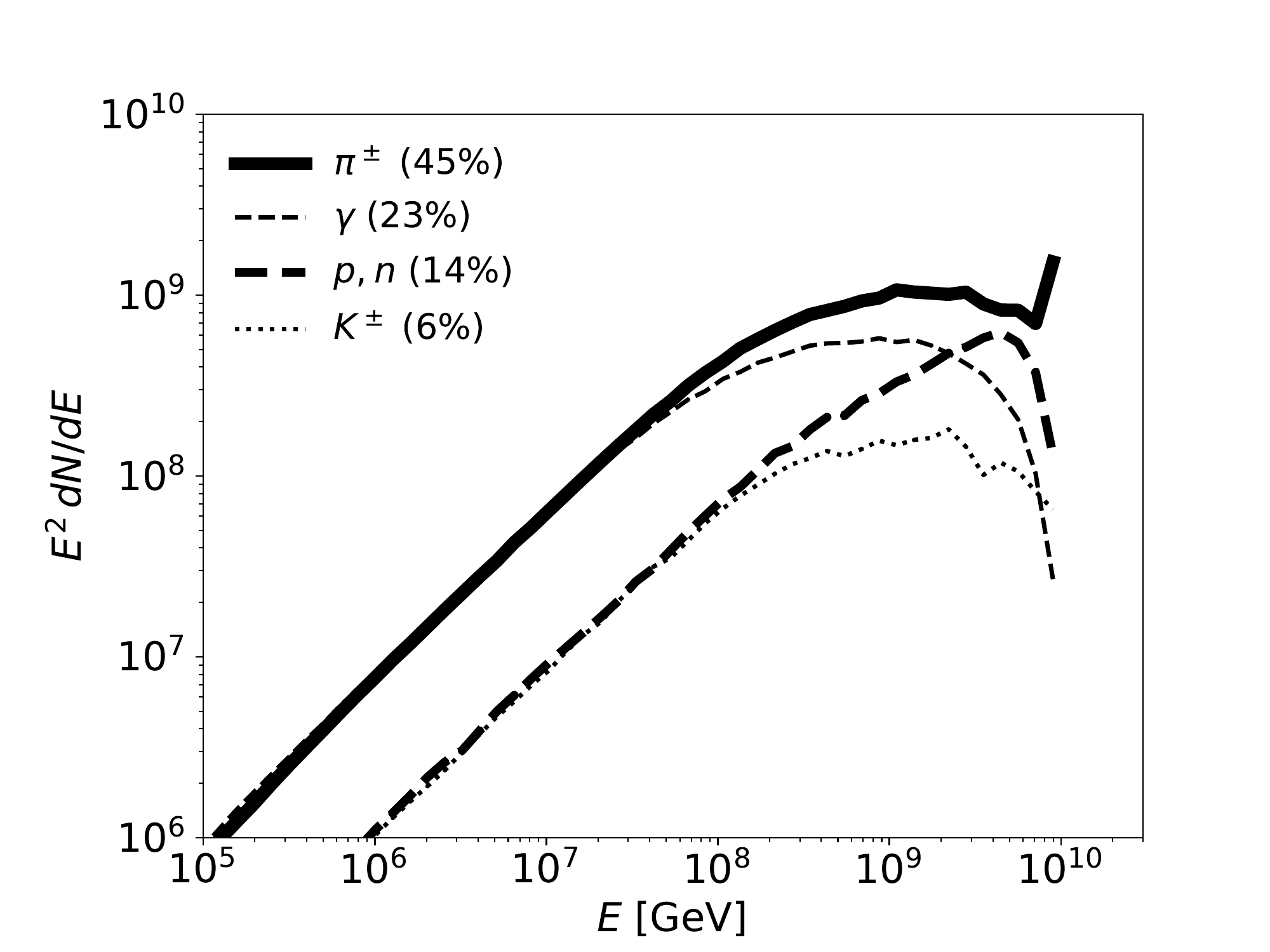}  
\caption{\label{fig:pip_productDist}
Distribution of products from the interaction of a pion of initial energy $10^{19}$~eV and a proton at rest, including charged pions $\pi^\pm$, gamma rays $\gamma$ from $\pi^0$ decays, kaons $K^\pm$, protons $p$, and neutrons $n$. The fraction of the energy carried by each group is indicated in the legend.
 }
\end{figure}

For reference, Figure~\ref{fig:pip_productDist} shows the distribution of outgoing particles from the interaction of an pion of energy $E=10^{19}$~eV with a proton at rest as calculated using the EPOS-LHC model. Roughly half of the energy of the primary pion is carried away by charged mesons, while $15\%$ is in the form of baryons.  These products will ultimately generate neutrinos.

To account for neutrinos produced by secondary pions, we record all charged meson and baryon products from the $pp$ and $p\gamma$ interaction. At any time, each of these intermediate products may undergo one of four processes, depending on the background densities at the current step: 1) cooling by $\pi p$ interaction; 2) cooling by synchrotron radiation; 3) decay into a muon and a muon neutrino (with a 100\% branching ratio for $\pi^\pm$ and a 63\% branching ratio for $K^\pm$); and 4) free propagation. The products of $\pi p$ interaction are computed using the EPOS-LHC model \citep{PhysRevC.92.034906}.  All secondaries and their higher-order products are tracked until either their energy falls below TeV (where the atmospheric background dominates over astrophysical sources), or they escape the source without further interaction.

\subsection{Neutrino Fluence of AT2018cow}
\begin{figure}[t!]
\includegraphics[width=  \linewidth] {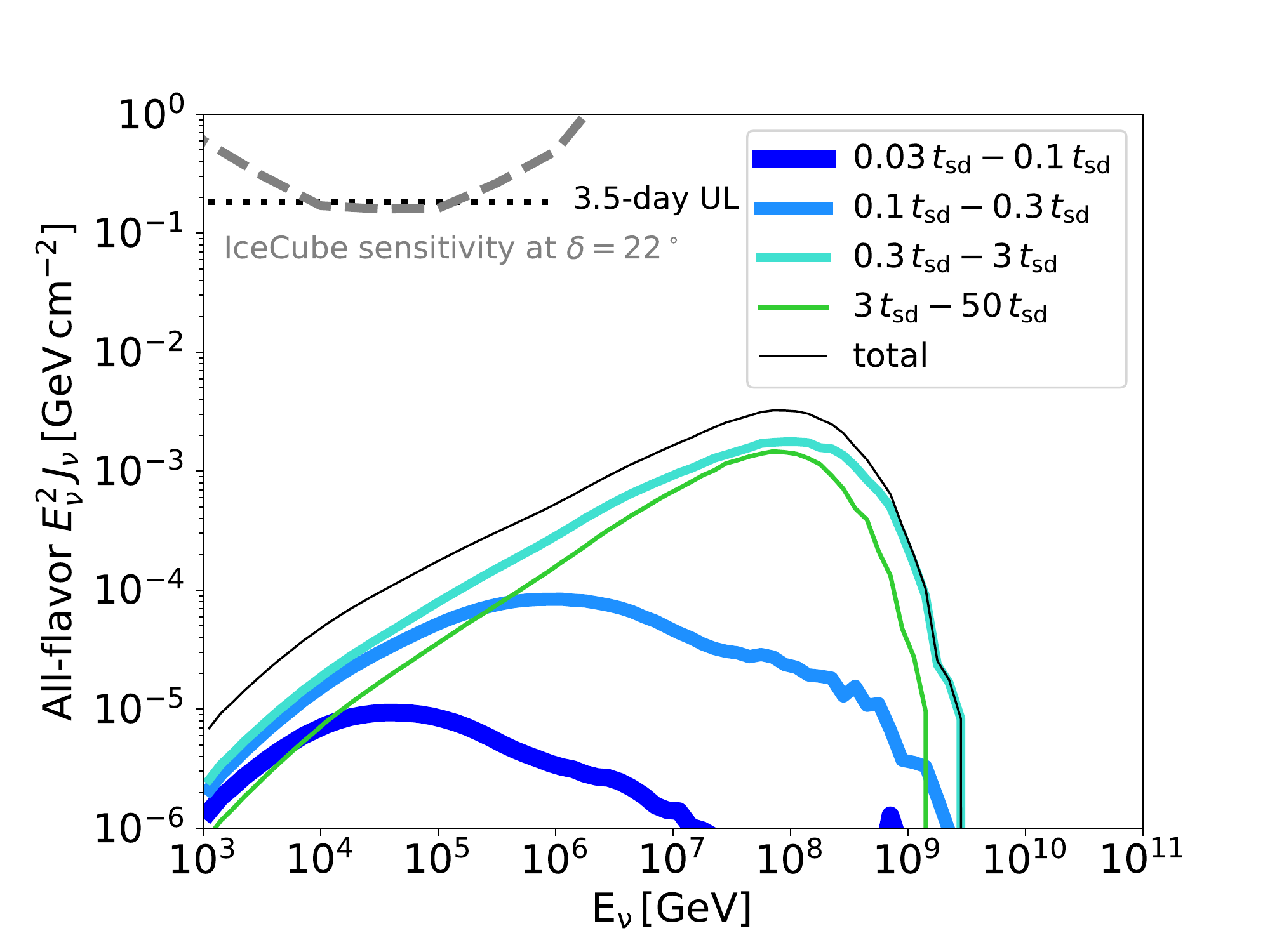}  
\includegraphics[width=  \linewidth] {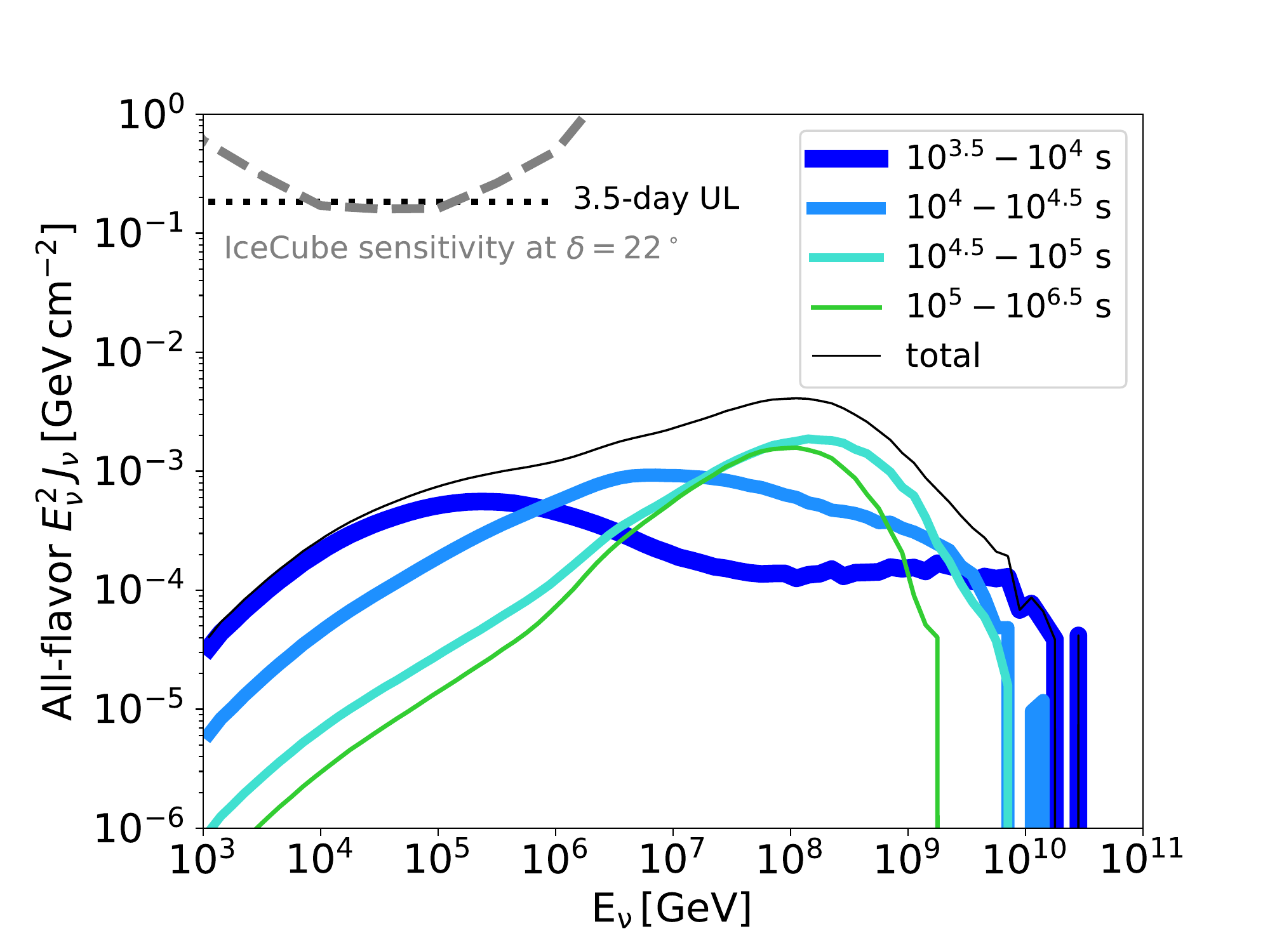}  
\caption{\label{fig:phi_neu}
All-flavor fluence of high-energy neutrinos from a magnetar-powered low-ejecta mass supernova on timescales ranging from one hour to $\sim$50 days after the explosion. For comparison, the time-integrated upper limit on $\nu_\mu$ flux $\left(E_\nu^2\,J_\nu\right)_{\rm UL}^{\rm IC}$ placed by the IceCube Observatory is shown (dotted line), which corresponds to a $1.8\,\sigma$ chance coincidence of neutrino events with the transient AT2018cow in 3.5 days. In addition, the IceCube sensitivity ($90\%$ C. L.) for point sources at $\delta=22^\circ$ is shown as a dashed curve, assuming an $E^{-2}$ spectrum over a half decade in energy \citep{2014ApJ...796..109A}. The parameters of the magnetar model are: 
initial spin period $P_i = 10$~ms, surface magnetic field $B_s = 10^{15}$~G (Case I; top), and $P_i = 2$~ms, $B_s = 2\times 10^{15}$~G (Case II; bottom).}
\end{figure}

\begin{figure}[t!] 
\includegraphics[width=  \linewidth] {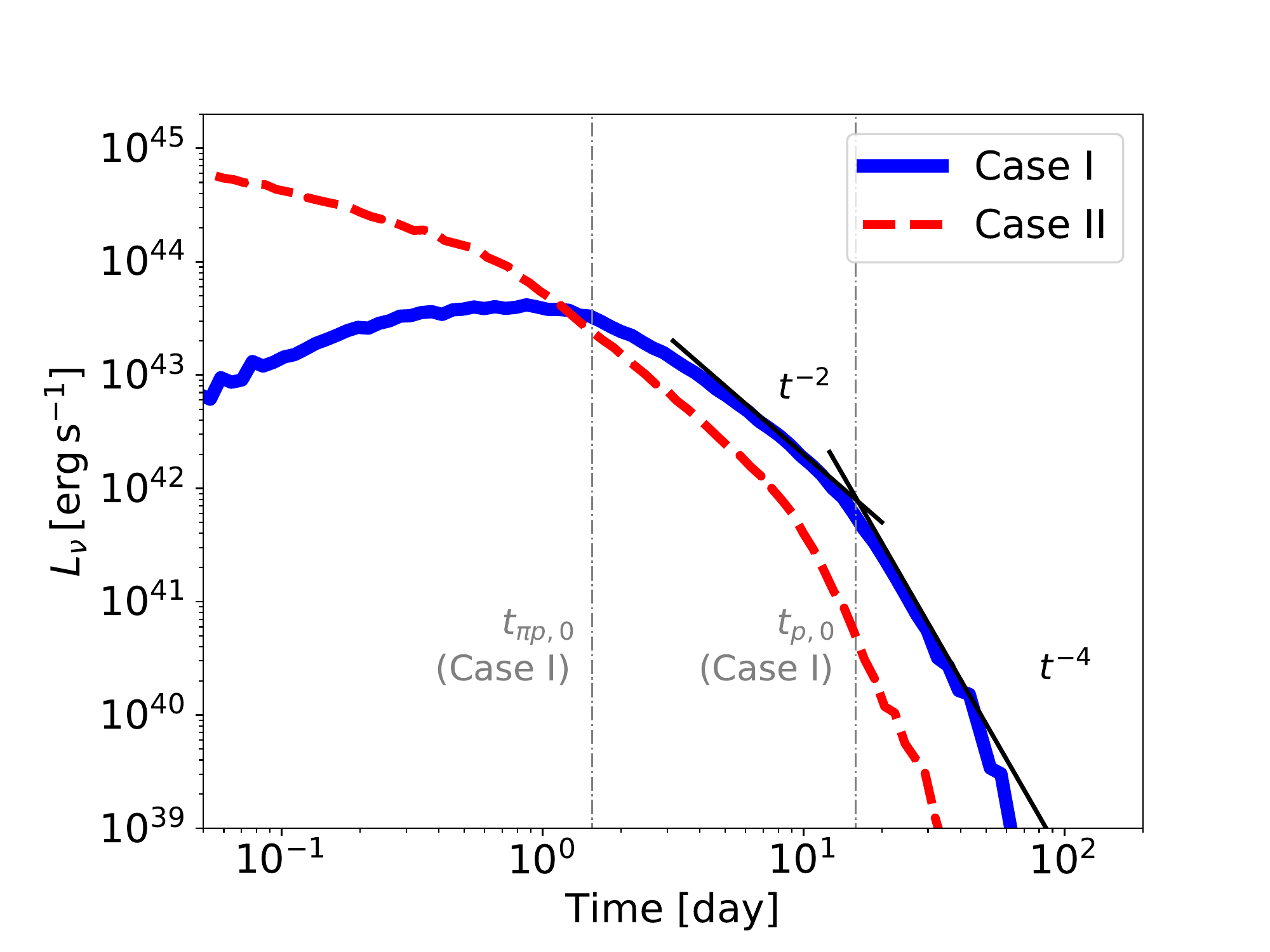}  
\caption{\label{fig:L_neu}
Luminosity of neutrinos with energies $> 1$ TeV as a function of time since explosion, shown separately for Case I (blue-solid line) and Case II (red-dashed line).  As indicated by black lines, the light curves approximately obey $L_\nu\propto t^{-2}$ at times $t_{\pi p,0} \lesssim t \lesssim t_{p,0}^{\rm ej}$, and $L_\nu \propto t^{-4}$ at times $t > t_{p,0}^{\rm ej}$.}
\end{figure}

Figure~\ref{fig:phi_neu} presents the neutrino fluence in the two fiducial models.  In Case I, we show results in time intervals normalized to the magnetar spin-down time $t_{\rm sd}$ (roughly one day). Neutrino production is low at both early times $t\ll t_{\rm sd}$, when the bulk of the cosmic rays have yet to be injected, and at late times $t_{\rm pp} > t_{\rm cross, ej}$, when $pp$ interactions become inefficient.  Most neutrinos are generated around the time $t_{\rm sd}$, when the magnetar releases most of its rotational energy and a sufficiently dense baryon background still exists for pion production. The turn-over in the neutrino spectrum is determined at early times by the break energy $E_{\nu,b}$ (eq.~\ref{eq:Enub}), while at late times the break is determined by the maximum cosmic ray energy $E_{\rm CR}$ (eq.~\ref{eqn:ECR}). The spectral index before the break is a convolution of the energy distribution of charged pions from the $pp$ and $p\gamma$ interaction (similar to that from $\pi p$ as shown in Figure~\ref{fig:pip_productDist}) and the history of particle injection, as described by equation~\ref{eqn:dNdE}. 

The bottom panel of Figure~\ref{fig:phi_neu} shows our results for Case II ($P_i = 2$~ms, $B_d =2\times 10^{15}$~G).  
The spin-down time in this case is much shorter $t_{\rm sd} = 0.01$~d, whereas the neutrino break energy exceeds 1 TeV only after times $t= 0.03\,M_{\rm ej, -0.5}^{1/3}\,\beta_{\rm ej, -1.3}^{-1}\,\rm d$. As a result, most cosmic rays are injected too early to generate neutrinos in the energy range of interest. As the magnetar releases most of its energy before the environment becomes optically-thin, significant TeV-PeV neutrinos are produced in the first 0.5~day due to the $\pi p$ interaction.  The neutrino spectrum at the earliest epoch features two peaks; the low-energy bump is from pion decay, while the tail at high energies arises from the decay of short-lived mesons other than $\pi^\pm$.  
After about one day, the ejecta becomes sufficiently dilute that meson cooling is no longer severe and the neutrino flux becomes maximal when the suppression is not important. Then, the neutrino spectrum comes to resemble that in Case I. This kind of time evolution of neutrino spectra owing to meson cooling in magnetars was first found in \cite{2009PhRvD..79j3001M} and \cite{2014PhRvD..90j3005F}. 

Figure~\ref{fig:L_neu} shows the total neutrino luminosity, $L_\nu=\int_{E_\nu > \rm TeV} \,E_\nu (dN_\nu/dE_\nu)dE_\nu dt$, at energies $>1$ TeV as a function of time since explosion.  In both Cases I and II the light curve obeys $L_\nu \propto t^{-2}$ at times $t_{\pi p,0}\lesssim t \lesssim t_{p,0}^{\rm ej}$, similar to the optical and X-ray light curves of AT2018cow (Fig.~\ref{fig:lightCurve}; e.g.~\citealt{CowX}).  In the time window when the ejecta is still optically thick to cosmic rays, yet after the pion cooling no longer suppresses neutrino production, the neutrino luminosity tracks the spin-down power of the magnetar, $L_{\rm sd} \propto t^{-2}$ (eq.~\ref{eqn:Lsd}). 

The origin of the time dependence of the neutrino luminosity is more complex at earlier and later times.  At early times $t\ll t_{\pi p, 0}$, both the cosmic ray injection rate $dN/dt$ and maximum energy $E_p$ decrease in time as $\left(1+t/t_{\rm sd}\right)^{-1}$, while the suppression factor due to $\pi$ cooling rises as $f_{\rm sup}\propto t^3 \left(1+t/t_{\rm sd}\right)$. If, as in Case II, the majority of the cosmic ray energy is injected at early times when the system is still optically thick to protons and pions, then the time evolution is also influenced by neutrinos released from $p\gamma$ interaction, $\pi p$ interaction and kaon decay. At late times $t\gg t_{p,0}^{\rm ej}$, the effective optical depth decreases as $\tau_{\nu} \propto t^{-2}$, and thus the neutrino light curve obeys a steeper decay, $L_{\nu} \sim L_{\rm sd}\tau_{\nu} \propto t^{-4}$~\citep[see also][]{2009PhRvD..79j3001M,2015JCAP...06..004F}. 

Our predictions for the neutrino fluence from AT2018cow is well below the upper limits placed by the IceCube Observatory for both models, supporting a conclusion that the two detected events are from the background rather than of astrophysical origin. A future event otherwise similar to AT2018cow but occurring $\sim 5$ times closer (at a distance of $\lesssim 15$ Mpc), would be a promising IceCube source. Prospects may be better for future neutrino telescopes with sensitivity in the EeV energy range, enabling a direct test of FBOTs as cosmic particle accelerators.

\subsection{Neutrino Flux of the SLSNe population}
\begin{figure}[t!] 
\includegraphics[width=  \linewidth] {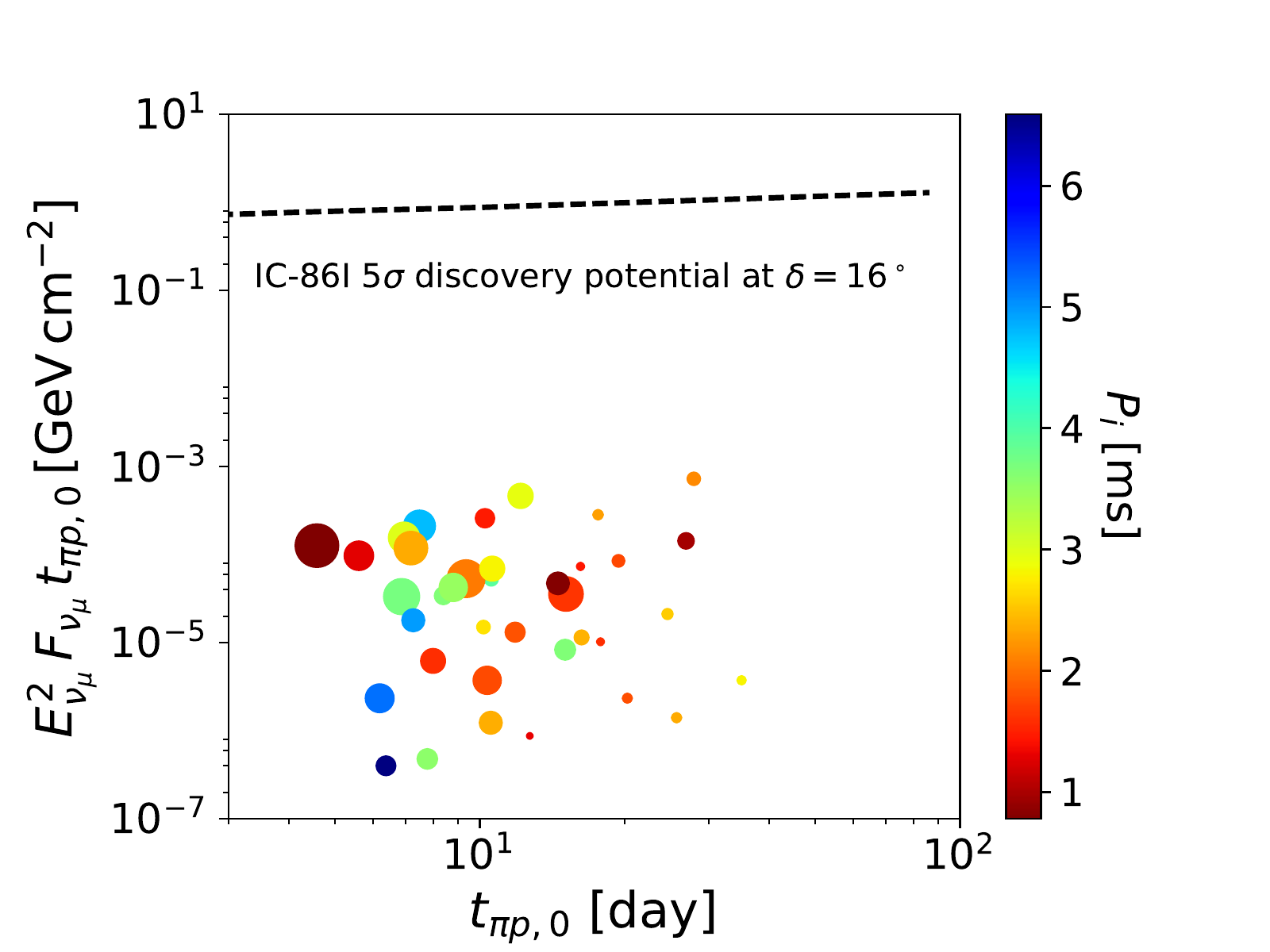}  
\caption{\label{fig:Fv_Nicholl}
Peak neutrino fluence as estimated by equation~\ref{eqn:F_nu} for each member of the sample of SLSNe with known distances and best-fit magnetar parameters from \citet{2017ApJ...850...55N}. The color of each symbol indicates the initial spin period while the symbol size indicates the strength of the magnetar's dipole magnetic field.  For reference, the dashed line shows the $5\sigma$ discovery potential of a time-dependent search of IceCube at declination $\delta =16^\circ$ with an $E^{-2}$ spectrum  \citep{2015ApJ...807...46A}.}
\end{figure}

The neutrino flux in our models peak around the time $t_{\pi p,0}$, after which pions have sufficient time to decay into neutrinos before cooling. 
We therefore can estimate the peak flux as
\beq
\left(E^2\,F_\nu\right)_{\rm pk}\approx \frac{1}{2}\,E_p(t_{\pi p, 0})\,\frac{dN_p}{dt}(t_{\pi p, 0})\,\frac{1}{4\pi D^2}
\eeq 
For a magnetar with $t_{\rm sd}\ll t_{\pi p,0}$ (as in our Case II), the peak flux is somewhat higher than this estimate due to additional contributions from secondaries.

Under the assumption that the peak neutrino flux lasts for a duration $\Delta t\sim t_{\pi p,0}$, the corresponding fluence can be estimated by 
\beq \label{eqn:F_nu}
\left(E^2\,J_\nu\right)_{\rm pk}\sim \left(E^2\,F_\nu\right)_{\rm pk}\,t_{\pi p,0}
\eeq
Figure~\ref{fig:Fv_Nicholl} shows the peak fluence for a sample of SLSNe \citep{2017ApJ...850...55N} estimated by equation~\ref{eqn:F_nu} using the magnetar parameters best-fit to the optical light curves and the distances of the sources. For comparison, we show the discovery potential of a time-dependent search by IceCube for a source at the declination $\delta=16^\circ$ \citep{2015ApJ...807...46A} for different source time integrations $\sim  t_{\pi p,0}$. The IceCube point-source sensitivity also depends on the declination of the source and is greatest for $\delta \approx 0$ \citep{2017ApJ...835..151A}. 

\subsection{Integrated Neutrino Flux of FBOT Population}

\begin{figure}[t!] 
\includegraphics[width=  \linewidth] {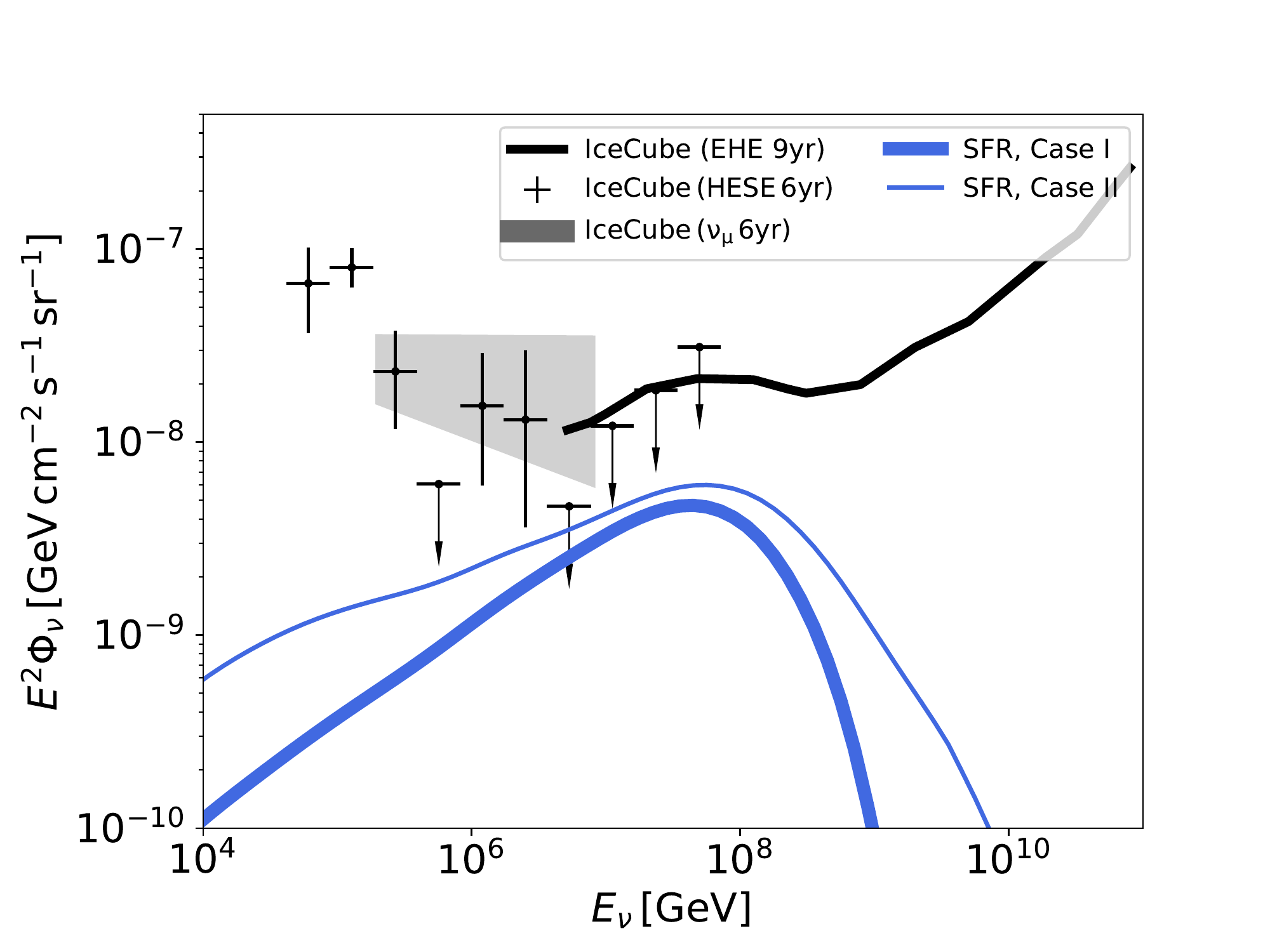}  
\caption{\label{fig:cumulativeFlux} 
Integrated neutrino flux over a cosmological source population of magnetar-powered FBOTs similar to AT2018cow, assuming an event rate of 1\% of the core-collapse supernovae rate in the local universe that evolves with redshift according the cosmic star formation rate \citep{0004-637X-651-1-142}. Results are shown separately for magnetar parameters corresponding to Case I and II, respectively. Flux constraints from the IceCube 6-year high-energy starting events (HESE) \citep{Kopper:2017zzm}, 6-year muon neutrino events \citep{2016ApJ...833....3A}, and 9-year extreme-high-energy (EHE) 90\% upper limit \citep{2018arXiv180701820I} are shown for comparison as labeled.} 
\end{figure}

The integrated neutrino flux over cosmological distances of magnetar-powered FBOTs with properties similar to AT2018cow
is given by
\beq
\Phi_\nu = \frac{c}{4\pi}\int_0^{z_{\rm max}} dz\,\rho(z) \,\left.\frac{dN}{dE'}\right|_{E'=(1+z)\,E}\,(1+z)\,\left|\frac{dt}{dz}\right|
\eeq
where $z$ is the source redshift, $|dz/dt| = H_0\,(1+z)\,\sqrt{\Omega_M\,(1+z)^3+\Omega_\Lambda}$ taking a flat $\Lambda$CDM with $\Omega_M=0.308$ and $H_0=67.8\,\rm km\,s^{-1}\,Mpc^{-1}$ \citep{2016A&A...594A..13P}. $\rho(z) =\rho_0 \,g(z)$ is the source birth rate at given redshift. We assume that AT2018cow-like FBOT events occur at $\sim 1\%$ of the core collapse supernova rate ($\rho_0 = 10^{-6}\,\rm Mpc^{-3}\,yr^{-1}$), and track the cosmological star-formation history  \citep{0004-637X-651-1-142} with $g(z) \propto (1+z)^{3.4}$ at $0<z<1$, then $g(z)\propto (1+z)^{-0.3}$ up to $1<z< 4$, and $(1+z)^{-3.5}$ at $z>4$.  
Figure~\ref{fig:cumulativeFlux} shows the integrated neutrino flux, calculated separately using neutrino fluences based on Case I and Case II, respectively. 
The peak flux is similar to that computed in \citet{2009PhRvD..79j3001M} after taking into account the difference in the source rates. This is because the peak fluence is determined by the time when pions start to decay rather than interact, and is thus insensitive to the factor $\beta_{\rm ej}^{3/4}\,M_{\rm ej}^{-1/4}$~\citep[see Equation~25 of this work and Equation~4 of][]{2009PhRvD..79j3001M}.
The flux is consistent with the extreme-high-energy upper limits from IceCube at 90\% C.L. \citep{2018arXiv180701820I}. Interestingly, if all FBOTs were as luminous as AT2018cow (i.e. if their rate was 4-7\% of the CCSN rate; \citealt{Drout+14}), then the integrated neutrino flux would have overproduced the IceCube limit.

\subsection{Gamma-Ray Emission from FBOTs and SLSNe}
\begin{figure}[t!] 
\includegraphics[width=  \linewidth] {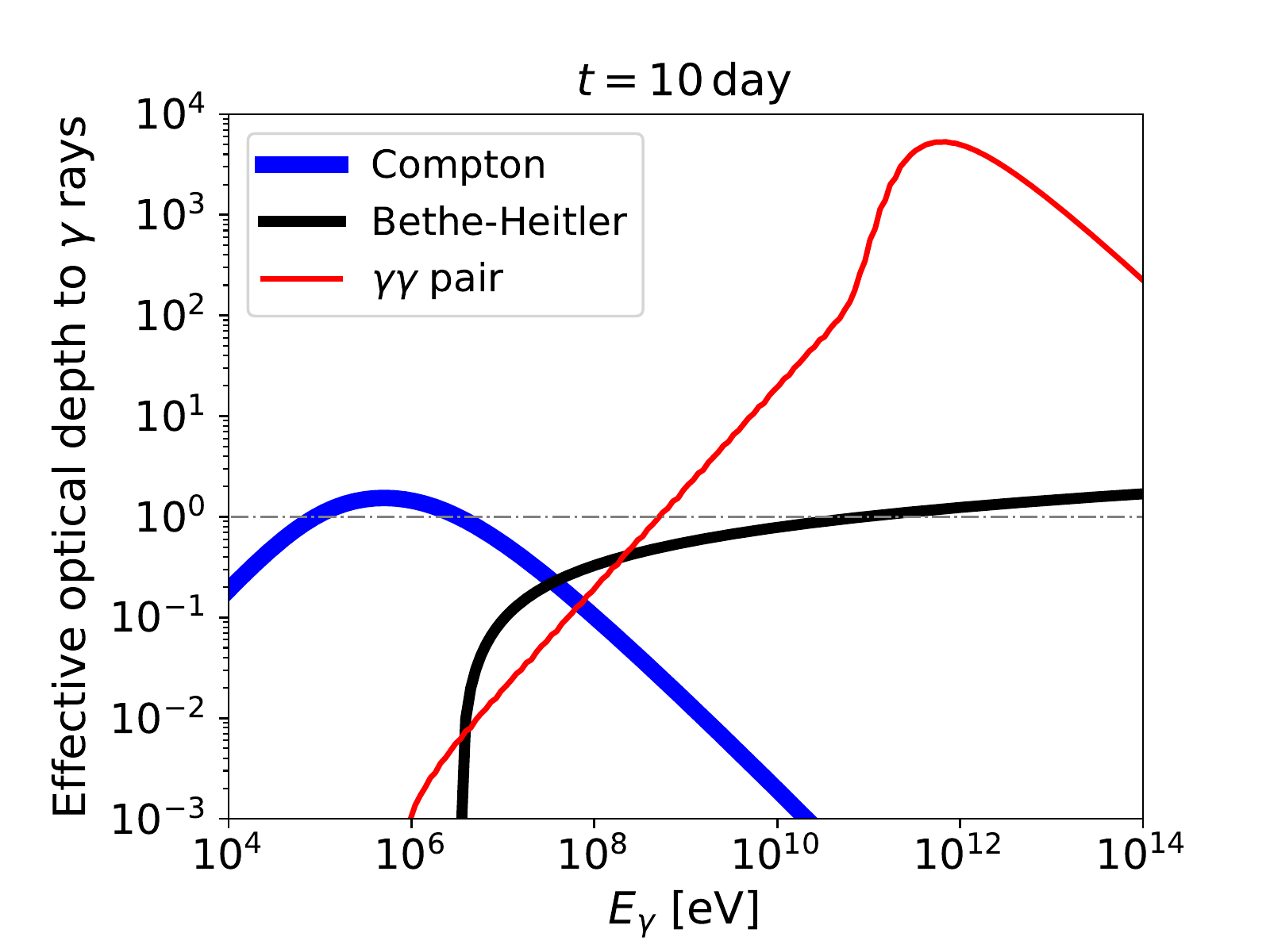}  
\caption{\label{fig:tau_gamma}
Optical depth to $\gamma$ rays at day 10. The dominant loss process for MeV, sub-GeV and high-energy $\gamma$ rays are Compton scattering, Bethe-Heitler pair production and $\gamma\gamma$ pair production respectively. At the plotting time, pair production of $\gamma$ rays from GeV to $0.1$~TeV is dominated by non-thermal photons from the nebula, and that above $0.1$~TeV is dominated by thermal photons from the ejecta.}
\end{figure}

Neutral pions created by UHE proton interactions produce high-energy $\gamma$ rays via $\pi^0\rightarrow 2\,\gamma$. In $pp$ and $p\gamma$ channels, the energy passed by protons to $\gamma$ rays is about $2/3$ and $4/3$ times that given to neutrinos, respectively.  High-energy gamma rays will quickly undergo pair production with low-energy photons in the nebula ($\gamma\gamma\rightarrow e^\pm$). 
For GeV to TeV $\gamma$ rays, the peak of the $\gamma\gamma$ pair production cross section lies in the optical to soft X-ray range. The optical depth of the background photon to high-energy $\gamma$ rays is approximately given by
\begin{eqnarray}
\tau_{\rm \gamma\gamma} &\sim& n_{\rm th}\,\sigma_{\gamma\gamma}\,R_{\rm ej} \nonumber \\
 &\approx& 2.0\times10^5\,P_{i,-2}^{-3/2}\,\beta_{\rm ej, -1.3}^{-1/2}\,\left(\frac{t}{1\,\rm day}\right)^{-2},
\end{eqnarray}
where $\sigma_{\gamma\gamma} = 3 / 16 \,\sigma_T$ is the $\gamma\gamma$ pair production cross section and $u_{\rm th}\sim L_{\rm sd} / \left(4\pi\,R_{\rm ej}^2\,c\right)$ is the thermal photon density at late times. Thus high-energy $\gamma$ rays can not escape from the ejecta freely until after a time 
\beq
t_{\gamma,0} \simeq 450\,P_{i,-2}^{-1}\,\beta_{\rm ej, -1.3}^{-1/2}\,\rm day
\eeq
The pair production and inverse Compton scattering of the resulting electrons leads to an electromagnetic cascade (e.g.~\citealt{Metzger+14,Metzger&Piro14,2015ApJ...805...82M}).  Figure~\ref{fig:tau_gamma} presents the optical depth of the nebula and ejecta to $\gamma$ rays at different energies around day 10. 
For MeV, sub-GeV, and $>$~GeV $\gamma$ rays, the dominate energy loss process is Compton scattering ($e\gamma \rightarrow e\gamma$), Bethe-Heitler pair production ($p\gamma \rightarrow p\gamma$) and $\gamma\gamma$ pair production respectively. The effective optical depth is defined as $\tau_{\rm eff} = Y_{\rm b}\,\sigma\,\kappa$, where $Y_{\rm b} = 3\,M_{\rm ej}/ \left(4\pi R_{\rm ej}^2 \, \mu_e\right)$ is the column density of the ejecta, and $\mu_e\approx 2$ is the mean molecular weight. The cross section $\sigma$ and inelasticity $\kappa$ of the interactions depend on the $\gamma$-ray energy (equations~40, 46, 48 of \citealt{2015ApJ...805...82M}, also see \citealt{2009herb.book.....D} and references therein). 
Figure~\ref{fig:tau_gamma} shows that comparing to high-energy $\gamma$ rays, MeV to sub-GeV $\gamma$ rays have a better chance to be observed at early times. 
Note that the optical depths due to photon-matter interactions are estimated assuming the spherical geometry. If X-rays from the engine are observed, a more complicated geometry seems necessary~\citep{CowX}, and then the escape of gamma rays is likely to be easier in such more realistic setups.  
Also, we remark that the optical depth to the two-photon annihilation process decreases as energy, so UHE photons can escape from the system.  

\citet{2015ApJ...805...82M} computed the $\gamma$-ray and hard X-ray emission from a magnetar-powered CCSN, taking into account details of electromagnetic cascades. 
Applying the numerical code to AT2018cow-like events to a low-mass ejecta with a magnetar, the GeV gamma-ray flux is estimated to be at the level of $10^{-12}\,\rm erg\,cm^{-2}\,s^{-1}$ at the distance of AT2018cow. 
If the ejecta is asymmetric, as suggested by X-ray observations, MeV-GeV $\gamma$ rays are likely to leak out from low-density regions in the similar geometry and be observed.

\citet{Renault18} searched for GeV $\gamma$ rays in the directions of a SLSNe sample with the Fermi-LAT data and found no signals. Assuming that SLSNe population are equally luminous and have an $E^{-2}$ spectrum, \citet{Renault18} concluded an upper limit at 95\% C. L. to the $1-10$~GeV luminosity $L_{\gamma, 1-10\,\rm GeV}<9.1\times10^{41}\,\rm erg\,s^{-1}$. Our scenario is consistent with this limit, as the total spin-down power is below $3\times10^{41}\,\rm erg\,s^{-1}$ after 100 days. 

For AT2018cow, a gamma-ray non-detection was reported by Fermi LAT.\footnote{http://www.astronomerstelegram.org/?read=11808} during an one-week interval from day 3 to day 10 since the epoch of detection. 
The flux upper limit at 10~day is approximately $\sim{10}^{-11}~{\rm erg}~{\rm cm}^{-2}~{\rm s}^{-1}$ \citep[cf.][]{Renault18,2018ApJ...854...60M}.    
HESS reported an 95\% C. L. upper limit of $5\times10^{-13}\,\rm cm^{-2}\,s^{-1}$ based their observation in the third week.\footnote{http://www.astronomerstelegram.org/?read=11956} 
Both are consistent with our model.  

\section{Discussion and Conclusions}\label{sec:discussion}

Rapidly-spinning magnetars have been proposed as the engines responsible for powering the optical light curves of superluminous supernovae.   The recent discovery of AT2018cow, the first local example of a superluminous transient with a low ejecta mass (a so-called FBOT), enabled the discovery of coincident time-variable X-ray emission consistent with the presence of a central engine, given that X-rays escape due to the asymmetry of the ejecta (e.g.~\citealt{CowATLAS,Perley+18,CowX,Ho+18}).  The engine behind AT2018cow  released more than $\sim 10^{50}$ ergs over a timescale of less than a few days, behind a low mass ejecta shell $M_{\rm ej} \lesssim 0.3M_{\odot}$.  If the engine is a millisecond magnetar spinning down in isolation (i.e. neglecting fall-back accretion), then fits to the light curve require an initial spin period of $\sim 2-10$~ms and a dipole magnetic field of $\sim 1-2\times 10^{15}$~G. 

For the same reasons they are capable of powerin large luminosities, the extreme magnetars responsible for SLSNe and FBOTs also possess strong electromagnetic potentials, making them potential sites for the acceleration of relativistic particles and even UHECRs. Events with particularly low ejecta masses, such as AT2018cow, provide a way to directly view X-rays from the magnetar nebula (through prompt photo-ionization of the ejecta shell; \citealt{Metzger+14}) and allow for the timely escape of accelerated cosmic rays relative to the bulk of the normal SLSNe population.  

Motivated by the discovery of AT2018cow as a local example of an FBOT which shows direct evidence for a central engine, we have calculated the interaction of high-energy cosmic rays accelerated near the magnetar with background particles of the nebula and ejecta.  We take into account the time-evolving thermal and non-thermal radiation field of the nebula, and track all primary and higher-order interaction products down to TeV energies. 

Our results are largely insensitive to the modeling of the background radiation fields. The photopion interaction is dominated by the optical/UV emission which is directly observed, and the proton-proton interaction depends on the density of baryons, which is well constrained by the rise time of the explosion and the spectroscopically observed ejecta velocity. Our results do, however, depend on whether the engine behind AT2018cow is truly a magnetar (e.g., as opposed to an internal shock from CSM interaction or an accreting black hole), and whether millisecond magnetars are indeed efficient particle accelerators.  
Nevertheless, the direct detection of time-variable X-rays from AT2018cow provides greater confidence in the central engine scenario than was available from previous SLSNe samples (which generally show no coincident X-ray emission, likely due to photoelectric attenuation by the larger ejecta shells; \citealt{CowX,Margalit+18}).

UHECR protons accelerated by millisecond magnetars in FBOTs and SLSNe escape the source with characteristic energies $\sim $ EeV. In addition, heavy elements may be synthesized efficiently in magnetar winds \citep{2011MNRAS.415.2495M} and destroyed inside nebulae \citep{2012ApJ...753...69H, 2014PhRvD..89d3012M}. 
Nuclei heavier than protons could tap $Z$ times more energy from the same electric potential. Their energy losses are dominated by photo-disintegration and hadronuclear interaction. \citet{FKO12} shows that they may escape from a massive ejecta of CCSNe with $E>10^{20}$~eV if the pulsar has a millisecond spin period and surface dipole field in the range $10^{12}-10^{13}$~G. Neutrino production by heavy nuclei depends on the interaction channel. It could be comparable to that by protons when hadronuclear interactions dominate \citep{2015JCAP...06..004F}, but lower when photo-disintegration dominates. We will leave a more detailed study to a future work. 
Engine-driven supernovae, including jet-driven ones, have been discussed as the sources of UHECRs, where particle acceleration sites have been attributed to internal shocks in outflows, external forward and reverse shocks~\citep[e.g.,][]{2008PhRvD..78b3005M,2007PhRvD..76h3009W,2011NatCo...2E.175C,2018PhRvD..97h3010Z}. Our physical model is different in the sense that UHECR acceleration occurs inside pulsar wind nebulae.

For AT2018cow, the cosmic ray energies that can be achieved are lower than for the bulk of the SLSNe population considered previously.  On the other hand, if relatively common events similar to AT2018cow provide the dominant UHECR source just above the ankle energy, a tail of rarer more powerful FBOTs or SLSNe (e.g.~those born with initial spin periods close to their minimum break-up value $P_0 \sim 1$ ms instead of $P_0 \gtrsim 2$ ms) could dominate the UHECR budget at the highest energies.

Magnetars with birth properties similar to those required to power AT2018cow may be in some ways optimal for neutrino production. 
This is because of the apparent coincidence that the magnetar spin-down timescale is comparable to the narrow time window within which the optical depth for cosmic ray interaction is still high, but the secondary pions are no longer efficiently cooled before decaying. 
FBOTs similar to AT2018cow can therefore be in principle ideal targets for neutrino telescopes.  

Depending on the uncertain volumetric rate of FBOTs giving birth to millisecond magnetars with properties similar to those required to explain AT2018cow, we find that such a population could explain $\sim 10-20\%$ of the IceCube astrophysical neutrino background. However, given the sensitivity of current-generation facilities, the detection of individual FBOTs will be challenging without the fortuitous discovery of a source located several times closer than AT2018cow (which, however, was itself already much nearer than the previous cosmological populations of FBOTs and SLSNe).
Nevertheless, IceCube neutrino events can be used to optimize the strategies of transient telescopes such as Zwicky Transient Facility for observing FBOTs.

The same interactions giving rise to neutrinos in FBOTs and SLSNe also inevitably give rise to high energy gamma-rays.  However, because of the high pair creation optical depth created by the thermal photons of the transient, high-energy gamma-rays require months to years to escape from the magnetar nebula and thus to become detectable from Earth. Nevertheless, well-studied optical light curves of some SLSNe show evidence for ``leakage" of engine energy at late times $\gtrsim$ months (e.g.~\citealt{Nicholl+18}), which is not observed to be escaping in the soft X-ray band \citep{CowX,Bhirombhakdi+18} and therefore may emerge first in the hard X-ray or gamma-ray band.  

Our work motivates FBOTs and SLSNe as potentially promising sources for future neutrino and $\gamma$-ray telescopes. The predicted neutrino spectrum peaks at $100$~PeV-EeV, which can be favorably probed by both EeV neutrino detectors such as Askaryan Radio Array  \citep{2012APh....35..457A}, ARIANNA  \citep{2017APh....90...50B}, GRAND \citep{2018arXiv181009994G} POEMMA \citep{2017ICRC...35..542O}, as well as TeV-PeV detectors such as KM3NeT \citep{MARGIOTTA201483} and IceCube-Gen2 \citep{2014arXiv1412.5106I}. Depending on the structure of the ejecta and the distance of the source, the $\gamma$ rays may be observed by existing wide-field telescopes such as Fermi and HAWC, and the next-generation Cherenkov Telescopes Array \citep{2017arXiv170905434C}. 

\bigskip
The authors thank Tanguy Pierog for helpful discussions on the pion-proton interaction calculations, Erik Blaufuss and the IceCube Collaboration for useful feedback. 
B.D.M.~is supported by NASA (grants HST-AR-15041.001-A, 80NSSC18K1708, a80NSSC17K0501).
K.M. is supported by the Alfred P. Sloan Foundation and NSF grant no. PHY-1620777. I.B. is grateful for the generous support of the University of Florida. K.K. is supported by the APACHE grant (ANR-16-CE31-0001) of the French Agence Nationale de la Recherche. 

\bibliography{CowNeutrino} 

\begin{thebibliography}{}
\expandafter\ifx\csname natexlab\endcsname\relax\def\natexlab#1{#1}\fi
\providecommand{\url}[1]{\href{#1}{#1}}

\bibitem[{{Aartsen} {et~al.}(2014){Aartsen}, {Ackermann}, {Adams}, {Aguilar},
  {Ahlers}, {Ahrens}, {Altmann}, {Anderson}, {Arguelles}, {Arlen},
  {Auffenberg}, {Bai}, {Barwick}, {Baum}, {Beatty}, {Becker Tjus}, {Becker},
  {BenZvi}, {Berghaus}, {Berley}, {Bernardini}, {Bernhard}, {Besson}, {Binder},
  {Bindig}, {Bissok}, {Blaufuss}, {Blumenthal}, {Boersma}, {Bohm}, {Bos},
  {Bose}, {B{\"o}ser}, {Botner}, {Brayeur}, {Bretz}, {Brown}, {Casey},
  {Casier}, {Cheung}, {Chirkin}, {Christov}, {Christy}, {Clark}, {Classen},
  {Clevermann}, {Coenders}, {Cowen}, {Cruz Silva}, {Danninger}, {Daughhetee},
  {Davis}, {Day}, {de Andr{\'e}}, {De Clercq}, {De Ridder}, {Desiati}, {de
  Vries}, {de With}, {DeYoung}, {D{\'\i}az-V{\'e}lez}, {Dunkman}, {Eagan},
  {Eberhardt}, {Eichmann}, {Eisch}, {Euler}, {Evenson}, {Fadiran}, {Fazely},
  {Fedynitch}, {Feintzeig}, {Felde}, {Feusels}, {Filimonov}, {Finley},
  {Fischer-Wasels}, {Flis}, {Franckowiak}, {Frantzen}, {Fuchs}, {Gaisser},
  {Gallagher}, {Gerhardt}, {Gier}, {Gladstone}, {Gl{\"u}senkamp},
  {Goldschmidt}, {Golup}, {Gonzalez}, {Goodman}, {G{\'o}ra}, {Grandmont},
  {Grant}, {Gretskov}, {Groh}, {Gro{\ss}}, {Ha}, {Haack}, {Haj Ismail},
  {Hallen}, {Hallgren}, {Halzen}, {Hanson}, {Hebecker}, {Heereman}, {Heinen},
  {Helbing}, {Hellauer}, {Hellwig}, {Hickford}, {Hill}, {Hoffman}, {Hoffmann},
  {Homeier}, {Hoshina}, {Huang}, {Huelsnitz}, {Hulth}, {Hultqvist}, {Hussain},
  {Ishihara}, {Jacobi}, {Jacobsen}, {Jagielski}, {Japaridze}, {Jero},
  {Jlelati}, {Jurkovic}, {Kaminsky}, {Kappes}, {Karg}, {Karle}, {Kauer},
  {Kelley}, {Kheirandish}, {Kiryluk}, {Kl{\"a}s}, {Klein}, {K{\"o}hne},
  {Kohnen}, {Kolanoski}, {Koob}, {K{\"o}pke}, {Kopper}, {Kopper}, {Koskinen},
  {Kowalski}, {Kriesten}, {Krings}, {Kroll}, {Kroll}, {Kunnen}, {Kurahashi},
  {Kuwabara}, {Labare}, {Larsen}, {Larson}, {Lesiak-Bzdak}, {Leuermann},
  {Leute}, {L{\"u}nemann}, {Mac{\'\i}as}, {Madsen}, {Maggi}, {Maruyama},
  {Mase}, {Matis}, {Maunu}, {McNally}, {Meagher}, {Medici}, {Meli}, {Meures},
  {Miarecki}, {Middell}, {Middlemas}, {Milke}, {Miller}, {Mohrmann},
  {Montaruli}, {Morse}, {Nahnhauer}, {Naumann}, {Niederhausen}, {Nowicki},
  {Nygren}, {Obertacke}, {Odrowski}, {Olivas}, {Omairat}, {O'Murchadha},
  {Palczewski}, {Paul}, {Penek}, {Pepper}, {P{\'e}rez de los Heros},
  {Pfendner}, {Pieloth}, {Pinat}, {Posselt}, {Price}, {Przybylski}, {P{\"u}tz},
  {Quinnan}, {R{\"a}del}, {Rameez}, {Rawlins}, {Redl}, {Rees}, {Reimann},
  {Resconi}, {Rhode}, {Richman}, {Riedel}, {Robertson}, {Rodrigues}, {Rongen},
  {Rott}, {Ruhe}, {Ruzybayev}, {Ryckbosch}, {Saba}, {Sander}, {Sandroos},
  {Santander}, {Sarkar}, {Schatto}, {Scheriau}, {Schmidt}, {Schmitz},
  {Schoenen}, {Sch{\"o}neberg}, {Sch{\"o}nwald}, {Schukraft}, {Schulte},
  {Schulz}, {Seckel}, {Sestayo}, {Seunarine}, {Shanidze}, {Sheremata}, {Smith},
  {Soldin}, {Spiczak}, {Spiering}, {Stamatikos}, {Stanev}, {Stanisha},
  {Stasik}, {Stezelberger}, {Stokstad}, {St{\"o}{\ss}l}, {Strahler},
  {Str{\"o}m}, {Strotjohann}, {Sullivan}, {Taavola}, {Taboada}, {Tamburro},
  {Tepe}, {Ter- Antonyan}, {Terliuk}, {Te{\v{s}}i{\'c}}, {Tilav}, {Toale},
  {Tobin}, {Tosi}, {Tselengidou}, {Unger}, {Usner}, {Vallecorsa}, {van
  Eijndhoven}, {Vandenbroucke}, {van Santen}, {Vehring}, {Voge}, {Vraeghe},
  {Walck}, {Wallraff}, {Weaver}, {Wellons}, {Wendt}, {Westerhoff}, {Whelan},
  {Whitehorn}, {Wichary}, {Wiebe}, {Wiebusch}, {Williams}, {Wissing}, {Wolf},
  {Wood}, {Woschnagg}, {Xu}, {Xu}, {Yanez}, {Yodh}, {Yoshida}, {Zarzhitsky},
  {Ziemann}, {Zierke}, {Zoll}, \& {IceCube
  Collaboration}}]{2014ApJ...796..109A}
{Aartsen}, M.~G., {Ackermann}, M., {Adams}, J., {et~al.} 2014, \apj, 796, 109

\bibitem[{{Aartsen} {et~al.}(2015){Aartsen}, {Ackermann}, {Adams}, {Aguilar},
  {Ahlers}, {Ahrens}, {Altmann}, {Anderson}, {Archinger}, {Arguelles}, {Arlen},
  {Auffenberg}, {Bai}, {Barwick}, {Baum}, {Bay}, {Baker}, {Beatty}, {Becker
  Tjus}, {Becker}, {BenZvi}, {Berghaus}, {Berley}, {Bernardini}, {Bernhard},
  {Besson}, {Binder}, {Bindig}, {Bissok}, {Blaufuss}, {Blumenthal}, {Boersma},
  {Bohm}, {Bos}, {Bose}, {B{\"o}ser}, {Botner}, {Brayeur}, {Bretz}, {Brown},
  {Buzinsky}, {Casey}, {Casier}, {Cheung}, {Chirkin}, {Christov}, {Christy},
  {Clark}, {Classen}, {Clevermann}, {Coenders}, {Cowen}, {Cruz Silva},
  {Daughhetee}, {Davis}, {Day}, {de Andr{\'e}}, {De Clercq}, {Dembinski}, {De
  Ridder}, {Desiati}, {de Vries}, {de Wasseige}, {de With}, {DeYoung},
  {D{\'\i}az-V{\'e}lez}, {Dumm}, {Dunkman}, {Eagan}, {Eberhardt}, {Ehrhardt},
  {Eichmann}, {Eisch}, {Euler}, {Evenson}, {Fadiran}, {Fazely}, {Fedynitch},
  {Feintzeig}, {Felde}, {Filimonov}, {Finley}, {Fischer-Wasels}, {Flis},
  {Frantzen}, {Fuchs}, {Gaisser}, {Gaior}, {Gallagher}, {Gerhardt}, {Gier},
  {Gladstone}, {Gl{\"u}senkamp}, {Goldschmidt}, {Golup}, {Gonzalez}, {Goodman},
  {G{\'o}ra}, {Grant}, {Gretskov}, {Groh}, {Gro{\ss}}, {Ha}, {Haack}, {Haj
  Ismail}, {Hallen}, {Hallgren}, {Halzen}, {Hanson}, {Hebecker}, {Heereman},
  {Heinen}, {Helbing}, {Hellauer}, {Hellwig}, {Hickford}, {Hignight}, {Hill},
  {Hoffman}, {Hoffmann}, {Homeier}, {Hoshina}, {Huang}, {Huelsnitz}, {Hulth},
  {Hultqvist}, {In}, {Ishihara}, {Jacobi}, {Jacobsen}, {Japaridze}, {Jero},
  {Jurkovic}, {Kaminsky}, {Kappes}, {Karg}, {Karle}, {Kauer}, {Keivani},
  {Kelley}, {Kheirandish}, {Kiryluk}, {Kl{\"a}s}, {Klein}, {K{\"o}hne},
  {Kohnen}, {Kolanoski}, {Koob}, {K{\"o}pke}, {Kopper}, {Kopper}, {Koskinen},
  {Kowalski}, {Krings}, {Kroll}, {Kroll}, {Kunnen}, {Kurahashi}, {Kuwabara},
  {Labare}, {Lanfranchi}, {Larsen}, {Larson}, {Lesiak-Bzdak}, {Leuermann},
  {L{\"u}nemann}, {Madsen}, {Maggi}, {Mahn}, {Maruyama}, {Mase}, {Matis},
  {Maunu}, {McNally}, {Meagher}, {Medici}, {Meli}, {Meures}, {Miarecki},
  {Middell}, {Middlemas}, {Milke}, {Miller}, {Mohrmann}, {Montaruli}, {Morse},
  {Nahnhauer}, {Naumann}, {Niederhausen}, {Nowicki}, {Nygren}, {Obertacke},
  {Olivas}, {Omairat}, {O'Murchadha}, {Palczewski}, {Paul}, {Pepper},
  {P{\'e}rez de los Heros}, {Pfendner}, {Pieloth}, {Pinat}, {Posselt}, {Price},
  {Przybylski}, {P{\"u}tz}, {Quinnan}, {R{\"a}del}, {Rameez}, {Rawlins},
  {Redl}, {Rees}, {Reimann}, {Relich}, {Resconi}, {Rhode}, {Richman}, {Riedel},
  {Robertson}, {Rodrigues}, {Rongen}, {Rott}, {Ruhe}, {Ruzybayev}, {Ryckbosch},
  {Saba}, {Sander}, {Sandroos}, {Santander}, {Sarkar}, {Schatto}, {Scheriau},
  {Schmidt}, {Schmitz}, {Schoenen}, {Sch{\"o}neberg}, {Sch{\"o}nwald},
  {Schukraft}, {Schulte}, {Schulz}, {Seckel}, {Sestayo}, {Seunarine},
  {Shanidze}, {Smith}, {Soldin}, {Spiczak}, {Spiering}, {Stamatikos}, {Stanev},
  {Stanisha}, {Stasik}, {Stezelberger}, {Stokstad}, {St{\"o}{\ss}l},
  {Strahler}, {Str{\"o}m}, {Strotjohann}, {Sullivan}, {Sutherland}, {Taavola},
  {Taboada}, {Tamburro}, {Ter-Antonyan}, {Terliuk}, {Te{\v{s}}i{\'c}}, {Tilav},
  {Toale}, {Tobin}, {Tosi}, {Tselengidou}, {Unger}, {Usner}, {Vallecorsa}, {van
  Eijndhoven}, {Vandenbroucke}, {van Santen}, {Vanheule}, {Vehring}, {Voge},
  {Vraeghe}, {Walck}, {Wallraff}, {Weaver}, {Wellons}, {Wendt}, {Westerhoff},
  {Whelan}, {Whitehorn}, {Wichary}, {Wiebe}, {Wiebusch}, {Williams}, {Wissing},
  {Wolf}, {Wood}, {Woschnagg}, {Xu}, {Xu}, {Xu}, {Yanez}, {Yodh}, {Yoshida},
  {Zarzhitsky}, {Ziemann}, {Zoll}, \& {IceCube
  Collaboration}}]{2015ApJ...807...46A}
---. 2015, \apj, 807, 46

\bibitem[{{Aartsen} {et~al.}(2016){Aartsen}, {Abraham}, {Ackermann}, {Adams},
  {Aguilar}, {Ahlers}, {Ahrens}, {Altmann}, {Andeen}, {Anderson}, {Ansseau},
  {Anton}, {Archinger}, {Arg{\"u}elles}, {Auffenberg}, {Axani}, {Bai},
  {Barwick}, {Baum}, {Bay}, {Beatty}, {Becker Tjus}, {Becker}, {BenZvi},
  {Berghaus}, {Berley}, {Bernardini}, {Bernhard}, {Besson}, {Binder}, {Bindig},
  {Bissok}, {Blaufuss}, {Blot}, {Bohm}, {B{\"o}rner}, {Bos}, {Bose},
  {B{\"o}ser}, {Botner}, {Braun}, {Brayeur}, {Bretz}, {Burgman}, {Carver},
  {Casier}, {Cheung}, {Chirkin}, {Christov}, {Clark}, {Classen}, {Coenders},
  {Collin}, {Conrad}, {Cowen}, {Cross}, {Day}, {de Andr{\'e}}, {De Clercq},
  {del Pino Rosendo}, {Dembinski}, {De Ridder}, {Desiati}, {de Vries}, {de
  Wasseige}, {de With}, {DeYoung}, {D{\'\i}az-V{\'e}lez}, {di Lorenzo},
  {Dujmovic}, {Dumm}, {Dunkman}, {Eberhardt}, {Ehrhardt}, {Eichmann}, {Eller},
  {Euler}, {Evenson}, {Fahey}, {Fazely}, {Feintzeig}, {Felde}, {Filimonov},
  {Finley}, {Flis}, {F{\"o}sig}, {Franckowiak}, {Friedman}, {Fuchs}, {Gaisser},
  {Gallagher}, {Gerhardt}, {Ghorbani}, {Giang}, {Gladstone}, {Glagla},
  {Gl{\"u}senkamp}, {Goldschmidt}, {Golup}, {Gonzalez}, {Grant}, {Griffith},
  {Haack}, {Haj Ismail}, {Hallgren}, {Halzen}, {Hansen}, {Hansmann},
  {Hansmann}, {Hanson}, {Hebecker}, {Heereman}, {Helbing}, {Hellauer},
  {Hickford}, {Hignight}, {Hill}, {Hoffman}, {Hoffmann}, {Holzapfel},
  {Hoshina}, {Huang}, {Huber}, {Hultqvist}, {In}, {Ishihara}, {Jacobi},
  {Japaridze}, {Jeong}, {Jero}, {Jones}, {Jurkovic}, {Kappes}, {Karg}, {Karle},
  {Katz}, {Kauer}, {Keivani}, {Kelley}, {Kemp}, {Kheirandish}, {Kim},
  {Kintscher}, {Kiryluk}, {Kittler}, {Klein}, {Kohnen}, {Koirala}, {Kolanoski},
  {Konietz}, {K{\"o}pke}, {Kopper}, {Kopper}, {Koskinen}, {Kowalski}, {Krings},
  {Kroll}, {Kr{\"u}ckl}, {Kr{\"u}ger}, {Kunnen}, {Kunwar}, {Kurahashi},
  {Kuwabara}, {Labare}, {Lanfranchi}, {Larson}, {Lauber}, {Lennarz}, {Lesiak-
  Bzdak}, {Leuermann}, {Leuner}, {Lu}, {L{\"u}nemann}, {Madsen}, {Maggi},
  {Mahn}, {Mancina}, {Mandelartz}, {Maruyama}, {Mase}, {Maunu}, {McNally},
  {Meagher}, {Medici}, {Meier}, {Meli}, {Menne}, {Merino}, {Meures},
  {Miarecki}, {Mohrmann}, {Montaruli}, {Moulai}, {Nahnhauer}, {Naumann},
  {Neer}, {Niederhausen}, {Nowicki}, {Nygren}, {Obertacke Pollmann}, {Olivas},
  {O'Murchadha}, {Palczewski}, {Pandya}, {Pankova}, {Peiffer}, {Penek},
  {Pepper}, {P{\'e}rez de los Heros}, {Pieloth}, {Pinat}, {Price},
  {Przybylski}, {Quinnan}, {Raab}, {R{\"a}del}, {Rameez}, {Rawlins}, {Reimann},
  {Relethford}, {Relich}, {Resconi}, {Rhode}, {Richman}, {Riedel}, {Robertson},
  {Rongen}, {Rott}, {Ruhe}, {Ryckbosch}, {Rysewyk}, {Sabbatini}, {Sanchez
  Herrera}, {Sandrock}, {Sandroos}, {Sarkar}, {Satalecka}, {Schimp},
  {Schlunder}, {Schmidt}, {Schoenen}, {Sch{\"o}neberg}, {Schumacher}, {Seckel},
  {Seunarine}, {Soldin}, {Song}, {Spiczak}, {Spiering}, {Stahlberg}, {Stanev},
  {Stasik}, {Steuer}, {Stezelberger}, {Stokstad}, {St{\"o}{\ss}l}, {Str{\"o}m},
  {Strotjohann}, {Sullivan}, {Sutherland}, {Taavola}, {Taboada}, {Tatar},
  {Tenholt}, {Ter- Antonyan}, {Terliuk}, {Te{\v{s}}i{\'c}}, {Tilav}, {Toale},
  {Tobin}, {Toscano}, {Tosi}, {Tselengidou}, {Turcati}, {Unger}, {Usner},
  {Vandenbroucke}, {van Eijndhoven}, {Vanheule}, {van Rossem}, {van Santen},
  {Veenkamp}, {Vehring}, {Voge}, {Vraeghe}, {Walck}, {Wallace}, {Wallraff},
  {Wandkowsky}, {Weaver}, {Weiss}, {Wendt}, {Westerhoff}, {Whelan}, {Wickmann},
  {Wiebe}, {Wiebusch}, {Wille}, {Williams}, {Wills}, {Wolf}, {Wood}, {Woolsey},
  {Woschnagg}, {Xu}, {Xu}, {Xu}, {Yanez}, {Yodh}, {Yoshida}, {Zoll}, \&
  {Icecube Collaboration}}]{2016ApJ...833....3A}
{Aartsen}, M.~G., {Abraham}, K., {Ackermann}, M., {et~al.} 2016, \apj, 833, 3

\bibitem[{{Aartsen} {et~al.}(2017){Aartsen}, {Abraham}, {Ackermann}, {Adams},
  {Aguilar}, {Ahlers}, {Ahrens}, {Altmann}, {Andeen}, {Anderson}, {Ansseau},
  {Anton}, {Archinger}, {Arg{\"u}elles}, {Auffenberg}, {Axani}, {Bai},
  {Barwick}, {Baum}, {Bay}, {Beatty}, {Becker Tjus}, {Becker}, {BenZvi},
  {Berley}, {Bernardini}, {Bernhard}, {Besson}, {Binder}, {Bindig}, {Bissok},
  {Blaufuss}, {Blot}, {Bohm}, {B{\"o}rner}, {Bos}, {Bose}, {B{\"o}ser},
  {Botner}, {Braun}, {Brayeur}, {Bretz}, {Bron}, {Burgman}, {Carver}, {Casier},
  {Cheung}, {Chirkin}, {Christov}, {Clark}, {Classen}, {Coenders}, {Collin},
  {Conrad}, {Cowen}, {Cross}, {Day}, {de Andr{\'e}}, {De Clercq}, {del Pino
  Rosendo}, {Dembinski}, {De Ridder}, {Desiati}, {de Vries}, {de Wasseige}, {de
  With}, {DeYoung}, {D{\'\i}az-V{\'e}lez}, {di Lorenzo}, {Dujmovic}, {Dumm},
  {Dunkman}, {Eberhardt}, {Ehrhardt}, {Eichmann}, {Eller}, {Euler}, {Evenson},
  {Fahey}, {Fazely}, {Feintzeig}, {Felde}, {Filimonov}, {Finley}, {Flis},
  {F{\"o}sig}, {Franckowiak}, {Friedman}, {Fuchs}, {Gaisser}, {Gallagher},
  {Gerhardt}, {Ghorbani}, {Giang}, {Gladstone}, {Glauch}, {Gl{\"u}senkamp},
  {Goldschmidt}, {Golup}, {Gonzalez}, {Grant}, {Griffith}, {Haack}, {Haj
  Ismail}, {Hallgren}, {Halzen}, {Hansen}, {Hansmann}, {Hanson}, {Hebecker},
  {Heereman}, {Helbing}, {Hellauer}, {Hickford}, {Hignight}, {Hill}, {Hoffman},
  {Hoffmann}, {Holzapfel}, {Hoshina}, {Huang}, {Huber}, {Hultqvist}, {In},
  {Ishihara}, {Jacobi}, {Japaridze}, {Jeong}, {Jero}, {Jones}, {Jurkovic},
  {Kappes}, {Karg}, {Karle}, {Katz}, {Kauer}, {Keivani}, {Kelley},
  {Kheirandish}, {Kim}, {Kintscher}, {Kiryluk}, {Kittler}, {Klein}, {Kohnen},
  {Koirala}, {Kolanoski}, {Konietz}, {K{\"o}pke}, {Kopper}, {Kopper},
  {Koskinen}, {Kowalski}, {Krings}, {Kroll}, {Kr{\"u}ckl}, {Kr{\"u}ger},
  {Kunnen}, {Kunwar}, {Kurahashi}, {Kuwabara}, {Labare}, {Lanfranchi},
  {Larson}, {Lauber}, {Lennarz}, {Lesiak-Bzdak}, {Leuermann}, {Lu},
  {L{\"u}nemann}, {Madsen}, {Maggi}, {Mahn}, {Mancina}, {Mandelartz},
  {Maruyama}, {Mase}, {Maunu}, {McNally}, {Meagher}, {Medici}, {Meier}, {Meli},
  {Menne}, {Merino}, {Meures}, {Miarecki}, {Mohrmann}, {Montaruli}, {Moulai},
  {Nahnhauer}, {Naumann}, {Neer}, {Niederhausen}, {Nowicki}, {Nygren},
  {Obertacke Pollmann}, {Olivas}, {O'Murchadha}, {Palczewski}, {Pandya},
  {Pankova}, {Peiffer}, {Penek}, {Pepper}, {P{\'e}rez de los Heros}, {Pieloth},
  {Pinat}, {Price}, {Przybylski}, {Quinnan}, {Raab}, {R{\"a}del}, {Rameez},
  {Rawlins}, {Reimann}, {Relethford}, {Relich}, {Resconi}, {Rhode}, {Richman},
  {Riedel}, {Robertson}, {Rongen}, {Rott}, {Ruhe}, {Ryckbosch}, {Rysewyk},
  {Sabbatini}, {Sanchez Herrera}, {Sandrock}, {Sandroos}, {Sarkar},
  {Satalecka}, {Schlunder}, {Schmidt}, {Schoenen}, {Sch{\"o}neberg},
  {Schumacher}, {Seckel}, {Seunarine}, {Soldin}, {Song}, {Spiczak}, {Spiering},
  {Stanev}, {Stasik}, {Stettner}, {Steuer}, {Stezelberger}, {Stokstad},
  {St{\"o}ssl}, {Str{\"o}m}, {Strotjohann}, {Sullivan}, {Sutherland},
  {Taavola}, {Taboada}, {Tatar}, {Tenholt}, {Ter- Antonyan}, {Terliuk},
  {Te{\v{s}}i{\'c}}, {Tilav}, {Toale}, {Tobin}, {Toscano}, {Tosi},
  {Tselengidou}, {Turcati}, {Unger}, {Usner}, {Vandenbroucke}, {van
  Eijndhoven}, {Vanheule}, {van Rossem}, {van Santen}, {Veenkamp}, {Vehring},
  {Voge}, {Vogel}, {Vraeghe}, {Walck}, {Wallace}, {Wallraff}, {Wandkowsky},
  {Weaver}, {Weiss}, {Wendt}, {Westerhoff}, {Whelan}, {Wickmann}, {Wiebe},
  {Wiebusch}, {Wille}, {Williams}, {Wills}, {Wolf}, {Wood}, {Woolsey},
  {Woschnagg}, {Xu}, {Xu}, {Xu}, {Yanez}, {Yodh}, {Yoshida}, {Zoll}, \&
  {<author pre=''(''>IceCube Collaboration}}]{2017ApJ...835..151A}
---. 2017, \apj, 835, 151

\bibitem[{Aartsen {et~al.}(2018)Aartsen, Ackermann, Adams, Aguilar, Ahlers,
  Ahrens, Al~Samarai, Altmann, Andeen, Anderson, Ansseau, Anton, Arg\"uelles,
  Auffenberg, Axani, Backes, Bagherpour, Bai, Barbano, Barron, Barwick, Baum,
  Bay, Beatty, Becker~Tjus, Becker, BenZvi, Berley, Bernardini, Besson, Binder,
  Bindig, Blaufuss, Blot, Bohm, B\"orner, Bos, B\"oser, Botner, Bourbeau,
  Bourbeau, Bradascio, Braun, Brenzke, Bretz, Bron, Brostean-Kaiser, Burgman,
  Busse, Carver, Cheung, Chirkin, Christov, Clark, Classen, Collin, Conrad,
  Coppin, Correa, Cowen, Cross, Dave, Day, de~Andr\'e, De~Clercq, DeLaunay,
  Dembinski, Deoskar, De~Ridder, Desiati, de~Vries, de~Wasseige, de~With,
  DeYoung, D\'{\i}az-V\'elez, di~Lorenzo, Dujmovic, Dumm, Dunkman, Dvorak,
  Eberhardt, Ehrhardt, Eichmann, Eller, Evenson, Fahey, Fazely, Felde,
  Filimonov, Finley, Flis, Franckowiak, Friedman, Fritz, Gaisser, Gallagher,
  Ganster, Gerhardt, Ghorbani, Giang, Glauch, Gl\"usenkamp, Goldschmidt,
  Gonzalez, Grant, Griffith, Haack, Hallgren, Halve, Halzen, Hanson, Hebecker,
  Heereman, Helbing, Hellauer, Hickford, Hignight, Hill, Hoffman, Hoffmann,
  Hoinka, Hokanson-Fasig, Hoshina, Huang, Huber, Hultqvist, H\"unnefeld,
  Hussain, In, Iovine, Ishihara, Jacobi, Japaridze, Jeong, Jero, Jones,
  Kalaczynski, Kang, Kappes, Kappesser, Karg, Karle, Katz, Kauer, Keivani,
  Kelley, Kheirandish, Kim, Kintscher, Kiryluk, Kittler, Klein, Koirala,
  Kolanoski, K\"opke, Kopper, Kopper, Koschinsky, Koskinen, Kowalski, Krings,
  Kroll, Kr\"uckl, Kunwar, Kurahashi, Kyriacou, Labare, Lanfranchi, Larson,
  Lauber, Leonard, Leuermann, Liu, Lohfink, Lozano~Mariscal, Lu, L\"unemann,
  Luszczak, Madsen, Maggi, Mahn, Makino, Mancina,
  Mari\ifmmode~\mbox{\c{s}}\else \c{s}\fi{}, Maruyama, Mase, Maunu, Meagher,
  Medici, Meier, Menne, Merino, Meures, Miarecki, Micallef, Moment\'e,
  Montaruli, Moore, Moulai, Nagai, Nahnhauer, Nakarmi, Naumann, Neer,
  Niederhausen, Nowicki, Nygren, Obertacke~Pollmann, Olivas, O'Murchadha,
  O'Sullivan, Palczewski, Pandya, Pankova, Peiffer, Pepper, P\'erez de~los
  Heros, Pieloth, Pinat, Pizzuto, Plum, Price, Przybylski, Raab, R\"adel,
  Rameez, Rauch, Rawlins, Rea, Reimann, Relethford, Renzi, Resconi, Rhode,
  Richman, Robertson, Rongen, Rott, Ruhe, Ryckbosch, Rysewyk, Safa,
  Sanchez~Herrera, Sandrock, Sandroos, Santander, Sarkar, Sarkar, Satalecka,
  Schaufel, Schlunder, Schmidt, Schneider, Schoenen, Sch\"oneberg, Schumacher,
  Sclafani, Seckel, Seunarine, Soedingrekso, Soldin, Song, Spiczak, Spiering,
  Stachurska, Stamatikos, Stanev, Stasik, Stein, Stettner, Steuer,
  Stezelberger, Stokstad, St\"o\ss{}l, Strotjohann, Stuttard, Sullivan,
  Sutherland, Taboada, Tenholt, Ter-Antonyan, Terliuk, Tilav, Toale, Tobin,
  T\"onnis, Toscano, Tosi, Tselengidou, Tung, Turcati, Turley, Ty, Unger,
  Usner, Vandenbroucke, Van~Driessche, van Eijk, van Eijndhoven, Vanheule, van
  Santen, Vraeghe, Walck, Wallace, Wallraff, Wandler, Wandkowsky, Watson, Waza,
  Weaver, Weiss, Wendt, Werthebach, Westerhoff, Whelan, Whitehorn, Wiebe,
  Wiebusch, Wille, Williams, Wills, Wolf, Wood, Wood, Woolsey, Woschnagg,
  Wrede, Xu, Xu, Xu, Yanez, Yodh, Yoshida, \& Yuan}]{2018arXiv180701820I}
Aartsen, M.~G., Ackermann, M., Adams, J., {et~al.} 2018, Phys. Rev. D, 98,
  062003.
\newblock \url{https://link.aps.org/doi/10.1103/PhysRevD.98.062003}

\bibitem[{{Alves Batista} {et~al.}(2016){Alves Batista}, {Dundovic}, {Erdmann},
  {Kampert}, {Kuempel}, {M{\"u}ller}, {Sigl}, {van Vliet}, {Walz}, \&
  {Winchen}}]{2016JCAP...05..038A}
{Alves Batista}, R., {Dundovic}, A., {Erdmann}, M., {et~al.} 2016, J. Cosmol.
  Astropart. Phys., 1605, 038

\bibitem[{{Ara Collaboration} {et~al.}(2012){Ara Collaboration}, {Allison},
  {Auffenberg}, {Bard}, {Beatty}, {Besson}, {B{\"o}ser}, {Chen}, {Chen},
  {Connolly}, {Davies}, {Duvernois}, {Fox}, {Gorham}, {Grashorn}, {Hanson},
  {Haugen}, {Helbing}, {Hill}, {Hoffman}, {Hong}, {Huang}, {Huang}, {Ishihara},
  {Karle}, {Kennedy}, {Landsman}, {Liu}, {Macchiarulo}, {Mase}, {Meures},
  {Meyhandan}, {Miki}, {Morse}, {Newcomb}, {Nichol}, {Ratzlaff}, {Richman},
  {Ritter}, {Rott}, {Rotter}, {Sandstrom}, {Seckel}, {Touart}, {Varner},
  {Wang}, {Weaver}, {Wendorff}, {Yoshida}, \& {Young}}]{2012APh....35..457A}
{Ara Collaboration}, {Allison}, P., {Auffenberg}, J., {et~al.} 2012,
  Astroparticle Physics, 35, 457

\bibitem[{{Arcavi} {et~al.}(2016){Arcavi}, {Wolf}, {Howell}, {Bildsten},
  {Leloudas}, {Hardin}, {Prajs}, {Perley}, {Svirski}, {Gal-Yam}, {Katz},
  {McCully}, {Cenko}, {Lidman}, {Sullivan}, {Valenti}, {Astier}, {Balland},
  {Carlberg}, {Conley}, {Fouchez}, {Guy}, {Pain}, {Palanque-Delabrouille},
  {Perrett}, {Pritchet}, {Regnault}, {Rich}, \& {Ruhlmann-Kleider}}]{Arcavi+16}
{Arcavi}, I., {Wolf}, W.~M., {Howell}, D.~A., {et~al.} 2016, \apj, 819, 35

\bibitem[{{Arons}(2003)}]{Arons03}
{Arons}, J. 2003, \apj, 589, 871

\bibitem[{{Barwick} {et~al.}(2017){Barwick}, {Besson}, {Burgman}, {Chiem},
  {Hallgren}, {Hanson}, {Klein}, {Kleinfelder}, {Nelles}, {Persichilli},
  {Phillips}, {Prakash}, {Reed}, {Shively}, {Tatar}, {Unger}, {Walker}, \&
  {Yodh}}]{2017APh....90...50B}
{Barwick}, S.~W., {Besson}, D.~Z., {Burgman}, A., {et~al.} 2017, Astroparticle
  Physics, 90, 50

\bibitem[{{Bhirombhakdi} {et~al.}(2018){Bhirombhakdi}, {Chornock}, {Margutti},
  {Nicholl}, {Metzger}, {Berger}, {Margalit}, \&
  {Milisavljevic}}]{Bhirombhakdi+18}
{Bhirombhakdi}, K., {Chornock}, R., {Margutti}, R., {et~al.} 2018, \apjl, 868,
  L32

\bibitem[{{Blasi} {et~al.}(2000){Blasi}, {Epstein}, \& {Olinto}}]{Blasi00}
{Blasi}, P., {Epstein}, R.~I., \& {Olinto}, A.~V. 2000, \apjl, 533, L123

\bibitem[{{Blaufuss}(2018)}]{Blaufuss18}
{Blaufuss}, E. 2018, The Astronomer's Telegram, 11785

\bibitem[{{Cerutti} \& {Beloborodov}(2017)}]{2017SSRv..207..111C}
{Cerutti}, B., \& {Beloborodov}, A.~M. 2017, \ssr, 207, 111

\bibitem[{{Chakraborti} {et~al.}(2011){Chakraborti}, {Ray}, {Soderberg},
  {Loeb}, \& {Chandra}}]{2011NatCo...2E.175C}
{Chakraborti}, S., {Ray}, A., {Soderberg}, A.~M., {Loeb}, A., \& {Chandra}, P.
  2011, Nature Communications, 2, 175

\bibitem[{{CTA Consortium}(2017)}]{2017arXiv170905434C}
{CTA Consortium}, T. 2017, arXiv e-prints, arXiv:1709.05434

\bibitem[{{Dall'Osso} {et~al.}(2009){Dall'Osso}, {Shore}, \&
  {Stella}}]{2009MNRAS.398.1869D}
{Dall'Osso}, S., {Shore}, S.~N., \& {Stella}, L. 2009, \mnras, 398, 1869

\bibitem[{{de Ugarte Postigo} {et~al.}(2018){de Ugarte Postigo}, {Bremer},
  {Kann}, {Izzo}, {Thoene}, {Schulze}, {Perley}, {Martin}, {Malesani},
  {Michalowski}, {de Gregorio-Monsalvo}, {Bensch}, {Blazek}, {Sanchez-Ramirez},
  {Kim}, \& {Krips}}]{deUgartePostigo+18}
{de Ugarte Postigo}, A., {Bremer}, M., {Kann}, D.~A., {et~al.} 2018, The
  Astronomer's Telegram, 11749

\bibitem[{{Dermer} \& {Menon}(2009)}]{2009herb.book.....D}
{Dermer}, C.~D., \& {Menon}, G. 2009, {High Energy Radiation from Black Holes:
  Gamma Rays, Cosmic Rays, and Neutrinos}

\bibitem[{{Drout} {et~al.}(2014){Drout}, {Chornock}, {Soderberg},
  {et~al.}}]{Drout+14}
{Drout}, M.~R., {Chornock}, R., {Soderberg}, A.~M., {et~al.} 2014, \apj, 794,
  23

\bibitem[{{Drout} {et~al.}(2013){Drout}, {Soderberg}, {Mazzali}, {Parrent},
  {Margutti}, {Milisavljevic}, {Sanders}, {Chornock}, {Foley}, {Kirshner},
  {Filippenko}, {Li}, {Brown}, {Cenko}, {Chakraborti}, {Challis}, {Friedman},
  {Ganeshalingam}, {Hicken}, {Jensen}, {Modjaz}, {Perets}, {Silverman}, \&
  {Wong}}]{Drout+13}
{Drout}, M.~R., {Soderberg}, A.~M., {Mazzali}, P.~A., {et~al.} 2013, \apj, 774,
  58

\bibitem[{{Eidelman} {et~al.}(2004){Eidelman}, {Hayes}, {Olive},
  {Aguilar-Benitez}, {Amsler}, {Asner}, {Babu}, {Barnett}, {Beringer},
  {Burchat}, {Carone}, {Caso}, {Conforto}, {Dahl}, {D'Ambrosio}, {Doser},
  {Feng}, {Gherghetta}, {Gibbons}, {Goodman}, {Grab}, {Groom}, {Gurtu},
  {Hagiwara}, {Hern{\'a}ndez-Rey}, {Hikasa}, {Honscheid}, {Jawahery}, {Kolda},
  {Kwon}, {Mangano}, {Manohar}, {March-Russell}, {Masoni}, {Miquel},
  {M{\"o}nig}, {Murayama}, {Nakamura}, {Navas}, {Pape}, {Patrignani}, {Piepke},
  {Raffelt}, {Roos}, {Tanabashi}, {Terning}, {T{\"o}rnqvist}, {Trippe},
  {Vogel}, {Wohl}, {Workman}, {Yao}, {Zyla}, {Armstrong}, {Gee}, {Harper},
  {Lugovsky}, {Lugovsky}, {Lugovsky}, {Rom}, {Artuso}, {Barberio}, {Battaglia},
  {Bichsel}, {Biebel}, {Bloch}, {Cahn}, {Casper}, {Cattai}, {Chivukula},
  {Cowan}, {Damour}, {Desler}, {Dobbs}, {Drees}, {Edwards}, {Edwards},
  {Elvira}, {Erler}, {Ezhela}, {Fetscher}, {Fields}, {Foster}, {Froidevaux},
  {Fukugita}, {Gaisser}, {Garren}, {Gerber}, {Gerbier}, {Gilman}, {Haber},
  {Hagmann}, {Hewett}, {Hinchliffe}, {Hogan}, {H{\"o}hler}, {Igo-Kemenes},
  {Jackson}, {Johnson}, {Karlen}, {Kayser}, {Kirkby}, {Klein}, {Kleinknecht},
  {Knowles}, {Kreitz}, {Kuyanov}, {Lahav}, {Langacker}, {Liddle}, {Littenberg},
  {Manley}, {Martin}, {Narain}, {Nason}, {Nir}, {Peacock}, {Quinn}, {Raby},
  {Ratcliff}, {Razuvaev}, {Renk}, {Rolandi}, {Ronan}, {Rosenberg}, {Sachrajda},
  {Sakai}, {Sanda}, {Sarkar}, {Schmitt}, {Schneider}, {Scott}, {Seligman},
  {Shaevitz}, {Sj{\"o}strand}, {Smoot}, {Spanier}, {Spieler}, {Spooner},
  {Srednicki}, {Stahl}, {Stanev}, {Suzuki}, {Tkachenko}, {Trilling},
  {Valencia}, {van Bibber}, {Vincter}, {Ward}, {Webber}, {Whalley},
  {Wolfenstein}, {Womersley}, {Woody}, {Zenin}, {Zhu}, \& {Particle Data
  Group}}]{2004PhLB..592....1E}
{Eidelman}, S., {Hayes}, K.~G., {Olive}, K.~A., {et~al.} 2004, Physics Letters
  B, 592, 1

\bibitem[{{Fang}(2015)}]{2015JCAP...06..004F}
{Fang}, K. 2015, \jcap, 6, 004

\bibitem[{{Fang} {et~al.}(2014){Fang}, {Kotera}, {Murase}, \&
  {Olinto}}]{2014PhRvD..90j3005F}
{Fang}, K., {Kotera}, K., {Murase}, K., \& {Olinto}, A.~V. 2014, \prd, 90,
  103005

\bibitem[{{Fang} {et~al.}(2016){Fang}, {Kotera}, {Murase}, \&
  {Olinto}}]{2016JCAP...04..010F}
---. 2016, \jcap, 4, 010

\bibitem[{{Fang} {et~al.}(2012){Fang}, {Kotera}, \& {Olinto}}]{FKO12}
{Fang}, K., {Kotera}, K., \& {Olinto}, A.~V. 2012, The Astrophysical Journal,
  750, 118

\bibitem[{{Fang} \& {Metzger}(2017)}]{Fang&Metzger17}
{Fang}, K., \& {Metzger}, B.~D. 2017, \apj, 849, 153

\bibitem[{{Fern{\'a}ndez} {et~al.}(2018){Fern{\'a}ndez}, {Quataert},
  {Kashiyama}, \& {Coughlin}}]{Fernandez+18}
{Fern{\'a}ndez}, R., {Quataert}, E., {Kashiyama}, K., \& {Coughlin}, E.~R.
  2018, \mnras, 476, 2366

\bibitem[{{GRAND Collaboration} {et~al.}(2018){GRAND Collaboration},
  {Alvarez-Muniz}, {Alves Batista}, {Balagopal V.}, {Bolmont}, {Bustamante},
  {Carvalho}, {Charrier}, {Cognard}, {Decoene}, {Denton}, {De Jong}, {De
  Vries}, {Engel}, {Fang}, {Finley}, {Gabici}, {Gou}, {Gu}, {Gu{\'e}pin}, {Hu},
  {Huang}, {Kotera}, {Le Coz}, {Lenain}, {Lu}, {Martineau-Huynh},
  {Mostaf{\'a}}, {Mottez}, {Murase}, {Niess}, {Oikonomou}, {Pierog}, {Qian},
  {Qin}, {Ran}, {Renault-Tinacci}, {Roth}, {Schr{\"o}der}, {Sch{\"u}ssler},
  {Tasse}, {Timmermans}, {Tueros}, {Wu}, {Zarka}, {Zech}, {Zhang}, {Zhang},
  {Zhang}, {Zheng}, \& {Zilles}}]{2018arXiv181009994G}
{GRAND Collaboration}, {Alvarez-Muniz}, J., {Alves Batista}, R., {et~al.} 2018,
  arXiv e-prints, arXiv:1810.09994

\bibitem[{{Ho} {et~al.}(2019){Ho}, {Phinney}, {Ravi}, {Kulkarni}, {Petitpas},
  {Emonts}, {Bhalerao}, {Blundell}, {Cenko}, {Dobie}, {Howie}, {Kamraj},
  {Kasliwal}, {Murphy}, {Perley}, {Sridharan}, \& {Yoon}}]{Ho+18}
{Ho}, A. Y.~Q., {Phinney}, E.~S., {Ravi}, V., {et~al.} 2019, \apj, 871, 73

\bibitem[{Hopkins \& Beacom(2006)}]{0004-637X-651-1-142}
Hopkins, A.~M., \& Beacom, J.~F. 2006, The Astrophysical Journal, 651, 142.
\newblock \url{http://stacks.iop.org/0004-637X/651/i=1/a=142}

\bibitem[{{Horiuchi} {et~al.}(2012){Horiuchi}, {Murase}, {Ioka}, \&
  {M{\'e}sz{\'a}ros}}]{2012ApJ...753...69H}
{Horiuchi}, S., {Murase}, K., {Ioka}, K., \& {M{\'e}sz{\'a}ros}, P. 2012, \apj,
  753, 69

\bibitem[{{Hotokezaka} {et~al.}(2017){Hotokezaka}, {Kashiyama}, \&
  {Murase}}]{2017ApJ...850...18H}
{Hotokezaka}, K., {Kashiyama}, K., \& {Murase}, K. 2017, \apj, 850, 18

\bibitem[{{IceCube-Gen2 Collaboration} {et~al.}(2014){IceCube-Gen2
  Collaboration}, {:}, {Aartsen}, {Ackermann}, {Adams}, {Aguilar}, {Ahlers},
  {Ahrens}, {Altmann}, {Anderson}, {Anton}, {Arguelles}, {Arlen}, {Auffenberg},
  {Axani}, {Bai}, {Bartos}, {Barwick}, {Baum}, {Bay}, {Beatty}, {Becker Tjus},
  {Becker}, {BenZvi}, {Berghaus}, {Berley}, {Bernardini}, {Bernhard}, {Besson},
  {Binder}, {Bindig}, {Bissok}, {Blaufuss}, {Blumenthal}, {Boersma}, {Bohm},
  {Bos}, {Bose}, {B{\"o}ser}, {Botner}, {Brayeur}, {Bretz}, {Brown},
  {Buzinsky}, {Casey}, {Casier}, {Cheung}, {Chirkin}, {Christov}, {Christy},
  {Clark}, {Classen}, {Clevermann}, {Coenders}, {Collin}, {Conrad}, {Cowen},
  {Cruz Silva}, {Daughhetee}, {Davis}, {Day}, {de Andr{\'e}}, {De Clercq}, {De
  Ridder}, {Desiati}, {de Vries}, {de With}, {DeYoung}, {andaz-V{\'e}lez},
  {Dunkman}, {Eagan}, {Eberhardt}, {Ehrhardt}, {Eichmann}, {Eisch}, {Euler},
  {Evans}, {Evenson}, {Fadiran}, {Fazely}, {Fedynitch}, {Feintzeig}, {Felde},
  {Filimonov}, {Finley}, {Fischer-Wasels}, {Flis}, {Frantzen}, {Fuchs},
  {Gaisser}, {Gaior}, {Gallagher}, {Gerhardt}, {Gier}, {Gladstone},
  {Gl{\"u}senkamp}, {Goldschmidt}, {Golup}, {Gonzalez}, {Goodman}, {G{\'o}ra},
  {Grant}, {Gretskov}, {Groh}, {Gro{\ss}}, {Ha}, {Haack}, {Haj Ismail},
  {Hallen}, {Hallgren}, {Halzen}, {Hanson}, {Haugen}, {Hebecker}, {Heereman},
  {Heinen}, {Helbing}, {Hellauer}, {Hellwig}, {Hickford}, {Hignight}, {Hill},
  {Hoffman}, {Hoffmann}, {Homeier}, {Hoshina}, {Huang}, {Huelsnitz}, {Hulth},
  {Hultqvist}, {Ishihara}, {Jacobi}, {Jacobsen}, {Japaridze}, {Jero},
  {Jlelati}, {Jones}, {Jurkovic}, {Kalekin}, {Kappes}, {Karg}, {Karle},
  {Katori}, {Katz}, {Kauer}, {Keivani}, {Kelley}, {Kheirandish}, {Kiryluk},
  {Kl{\"a}s}, {Klein}, {K{\"o}hne}, {Kohnen}, {Kolanoski}, {Koob}, {K{\"o}pke},
  {Kopper}, {Kopper}, {Koskinen}, {Kowalski}, {Krauss}, {Kriesten}, {Krings},
  {Kroll}, {Kroll}, {Kunnen}, {Kurahashi}, {Kuwabara}, {Labare}, {Lanfranchi},
  {Larsen}, {Larson}, {Lesiak-Bzdak}, {Leuermann}, {LoSecco}, {L{\"u}nemann},
  {Madsen}, {Maggi}, {Mahn}, {Marka}, {Marka}, {Maruyama}, {Mase}, {Matis},
  {Maunu}, {McNally}, {Meagher}, {Medici}, {Meli}, {Meures}, {Miarecki},
  {Middell}, {Middlemas}, {Milke}, {Miller}, {Mohrmann}, {Montaruli}, {Moore},
  {Morse}, {Nahnhauer}, {Naumann}, {Niederhausen}, {Nowicki}, {Nygren},
  {Obertacke}, {Odrowski}, {Olivas}, {Omairat}, {O'Murchadha}, {Palczewski},
  {Paul}, {Penek}, {Pepper}, {P{\'e}rez de los Heros}, {Pfendner}, {Pieloth},
  {Pinat}, {Pinfold}, {Posselt}, {Price}, {Przybylski}, {P{\"u}tz}, {Quinnan},
  {R{\"a}del}, {Rameez}, {Rawlins}, {Redl}, {Rees}, {Reimann}, {Relich},
  {Resconi}, {Rhode}, {Richman}, {Riedel}, {Robertson}, {Rodrigues}, {Rongen},
  {Rott}, {Ruhe}, {Ruzybayev}, {Ryckbosch}, {Saba}, {Sander}, {Sandroos},
  {Sandstrom}, {Santander}, {Sarkar}, {Schatto}, {Scheriau}, {Schmidt},
  {Schmitz}, {Schoenen}, {Sch{\"o}neberg}, {Sch{\"o}nwald}, {Schukraft},
  {Schulte}, {Schulz}, {Seckel}, {Sestayo}, {Seunarine}, {Shaevitz},
  {Shanidze}, {Smith}, {Soldin}, {S{\"o}ldner-Rembold}, {Spiczak}, {Spiering},
  {Stamatikos}, {Stanev}, {Stanisha}, {Stasik}, {Stezelberger}, {Stokstad},
  {St{\"o}{\ss} andl}, {Strahler}, {Str{\"o}m}, {Strotjohann}, {Sullivan},
  {Taavola}, {Taboada}, {Taketa}, {Tamburro}, {Tanaka}, {Tepe}, {Ter-Antonyan},
  {Terliuk}, {Te{\v{s}}}, {andi{\'c}}, {Tilav}, {Toale}, {Tobin}, {Tosi},
  {Tselengidou}, {Unger}, {Usner}, {Vallecorsa}, {van Eijndhoven},
  {Vandenbroucke}, {van Santen}, {Vanheule}, {Vehring}, {Voge}, {Vraeghe},
  {Walck}, {Wallraff}, {Weaver}, {Wellons}, {Wendt}, {Westerhoff}, {Whelan},
  {Whitehorn}, {Wichary}, {Wiebe}, {Wiebusch}, {Williams}, {Wissing}, {Wolf},
  {Wood}, {Woschnagg}, {Wren}, {Xu}, {Xu}, {Xu}, {Yanez}, {Yodh}, {Yoshida},
  {Zarzhitsky}, {Ziemann}, \& {Zoll}}]{2014arXiv1412.5106I}
{IceCube-Gen2 Collaboration}, {:}, {Aartsen}, M.~G., {et~al.} 2014, arXiv
  e-prints, arXiv:1412.5106

\bibitem[{{Iwamoto} {et~al.}(2017){Iwamoto}, {Amano}, {Hoshino}, \&
  {Matsumoto}}]{2017ApJ...840...52I}
{Iwamoto}, M., {Amano}, T., {Hoshino}, M., \& {Matsumoto}, Y. 2017, \apj, 840,
  52

\bibitem[{{Kasen} \& {Bildsten}(2010)}]{Kasen&Bildsten10}
{Kasen}, D., \& {Bildsten}, L. 2010, Astrophys. J., 717, 245

\bibitem[{{Kashiyama} {et~al.}(2016){Kashiyama}, {Murase}, {Bartos}, {Kiuchi},
  \& {Margutti}}]{2016ApJ...818...94K}
{Kashiyama}, K., {Murase}, K., {Bartos}, I., {Kiuchi}, K., \& {Margutti}, R.
  2016, \apj, 818, 94

\bibitem[{{Kennel} \& {Coroniti}(1984)}]{1984ApJ...283..694K}
{Kennel}, C.~F., \& {Coroniti}, F.~V. 1984, \apj, 283, 710

\bibitem[{{Kleiser} {et~al.}(2018){Kleiser}, {Kasen}, \&
  {Duffell}}]{Kleiser&Kasen18}
{Kleiser}, I.~K.~W., {Kasen}, D., \& {Duffell}, P.~C. 2018, \mnras, 475, 3152

\bibitem[{Kopper(2018)}]{Kopper:2017zzm}
Kopper, C. 2018, PoS, ICRC2017, 981

\bibitem[{{Kotera}(2011)}]{Kotera2011}
{Kotera}, K. 2011, \prd, 84, 023002

\bibitem[{Kotera {et~al.}(2015)Kotera, Amato, \& Blasi}]{Kotera_2015}
Kotera, K., Amato, E., \& Blasi, P. 2015, Journal of Cosmology and
  Astroparticle Physics, 2015, 026.
\newblock \url{https://doi.org/10.1088%2F1475-7516%2F2015%2F08%2F026}

\bibitem[{{Kotera} {et~al.}(2013){Kotera}, {Phinney}, \& {Olinto}}]{Kotera+13}
{Kotera}, K., {Phinney}, E.~S., \& {Olinto}, A.~V. 2013, \mnras, 432, 3228

\bibitem[{{Kuin} {et~al.}(2019){Kuin}, {Wu}, {Oates}, {Lien}, {Emery},
  {Kennea}, {de Pasquale}, {Han}, {Brown}, {Tohuvavohu}, {Breeveld}, {Burrows},
  {Cenko}, {Campana}, {Levan}, {Markwardt}, {Osborne}, {Page}, {Page},
  {Sbarufatti}, {Siegel}, \& {Troja}}]{Kuin+18}
{Kuin}, N.~{Paul}, M., {Wu}, K., {Oates}, S., {et~al.} 2019, \mnras, 53

\bibitem[{Liu {et~al.}(2018)Liu, Zhang, Wang, \& Dai}]{2041-8205-868-2-L24}
Liu, L.-D., Zhang, B., Wang, L.-J., \& Dai, Z.-G. 2018, The Astrophysical
  Journal Letters, 868, L24.
\newblock \url{http://stacks.iop.org/2041-8205/868/i=2/a=L24}

\bibitem[{{Lyutikov} \& {Toonen}(2018)}]{2018arXiv181207569L}
{Lyutikov}, M., \& {Toonen}, S. 2018, arXiv e-prints, arXiv:1812.07569

\bibitem[{{Margalit} {et~al.}(2018){Margalit}, {Metzger}, {Berger}, {Nicholl},
  {Eftekhari}, \& {Margutti}}]{Margalit+18}
{Margalit}, B., {Metzger}, B.~D., {Berger}, E., {et~al.} 2018, \mnras, 481,
  2407

\bibitem[{Margiotta(2014)}]{MARGIOTTA201483}
Margiotta, A. 2014, Nuclear Instruments and Methods in Physics Research Section
  A: Accelerators, Spectrometers, Detectors and Associated Equipment, 766, 83 ,
  rICH2013 Proceedings of the Eighth International Workshop on Ring Imaging
  Cherenkov Detectors Shonan, Kanagawa, Japan, December 2-6, 2013.
\newblock
  \url{http://www.sciencedirect.com/science/article/pii/S0168900214006433}

\bibitem[{{Margutti} {et~al.}(2018){Margutti}, {Metzger}, {Chornock}, {Vurm},
  {Roth}, {Grefenstette}, {Savchenko}, {Cartier}, {Steiner}, {Terreran},
  {Migliori}, {Milisavljevic}, {Alexander}, {Bietenholz}, {Blanchard}, {Bozzo},
  {Brethauer}, {Chilingarian}, {Coppejans}, {Ducci}, {Ferrigno}, {Fong},
  {G{\"O}tz}, {Guidorzi}, {Hajela}, {Hurley}, {Kuulkers}, {Laurent},
  {Mereghetti}, {Nicholl}, {Patnaude}, {Ubertini}, {Banovetz}, {Bartel},
  {Berger}, {Coughlin}, {Eftekhari}, {Frederiks}, {Kozlova}, {Laskar},
  {Svinkin}, {Drout}, {Macfadyen}, \& {Paterson}}]{CowX}
{Margutti}, R., {Metzger}, B.~D., {Chornock}, R., {et~al.} 2018, ArXiv
  e-prints, arXiv:1810.10720

\bibitem[{Matthews(2018)}]{Matthews:2017waf}
Matthews, J. 2018, PoS, ICRC2017, 1096

\bibitem[{Metzger {et~al.}(2018)Metzger, Beniamini, \&
  Giannios}]{0004-637X-857-2-95}
Metzger, B.~D., Beniamini, P., \& Giannios, D. 2018, The Astrophysical Journal,
  857, 95.
\newblock \url{http://stacks.iop.org/0004-637X/857/i=2/a=95}

\bibitem[{{Metzger} {et~al.}(2011){Metzger}, {Giannios}, \&
  {Horiuchi}}]{2011MNRAS.415.2495M}
{Metzger}, B.~D., {Giannios}, D., \& {Horiuchi}, S. 2011, \mnras, 415, 2495

\bibitem[{{Metzger} \& {Piro}(2014)}]{Metzger&Piro14}
{Metzger}, B.~D., \& {Piro}, A.~L. 2014, Mon. Not. R. Astron. Soc., 439, 3916

\bibitem[{{Metzger} {et~al.}(2008){Metzger}, {Quataert}, \&
  {Thompson}}]{Metzger+08}
{Metzger}, B.~D., {Quataert}, E., \& {Thompson}, T.~A. 2008, \mnras, 385, 1455

\bibitem[{{Metzger} {et~al.}(2014){Metzger}, {Vurm}, {Hasco{\"e}t}, \&
  {Beloborodov}}]{Metzger+14}
{Metzger}, B.~D., {Vurm}, I., {Hasco{\"e}t}, R., \& {Beloborodov}, A.~M. 2014,
  Mon. Not. R. Astron. Soc., 437, 703

\bibitem[{{Moriya} \& {Eldridge}(2016)}]{Moriya&Eldridge16}
{Moriya}, T.~J., \& {Eldridge}, J.~J. 2016, \mnras, 461, 2155

\bibitem[{{M{\"u}cke} {et~al.}(2000){M{\"u}cke}, {Engel}, {Rachen},
  {Protheroe}, \& {Stanev}}]{2000CoPhC.124..290M}
{M{\"u}cke}, A., {Engel}, R., {Rachen}, J.~P., {Protheroe}, R.~J., \& {Stanev},
  T. 2000, Computer Physics Communications, 124, 290

\bibitem[{{Murase} {et~al.}(2014){Murase}, {Dasgupta}, \&
  {Thompson}}]{2014PhRvD..89d3012M}
{Murase}, K., {Dasgupta}, B., \& {Thompson}, T.~A. 2014, \prd, 89, 043012

\bibitem[{{Murase} \& {Fukugita}(2019)}]{2018arXiv180604194M}
{Murase}, K., \& {Fukugita}, M. 2019, \prd, 99, 063012

\bibitem[{{Murase} {et~al.}(2008){Murase}, {Ioka}, {Nagataki}, \&
  {Nakamura}}]{2008PhRvD..78b3005M}
{Murase}, K., {Ioka}, K., {Nagataki}, S., \& {Nakamura}, T. 2008, \prd, 78,
  023005

\bibitem[{{Murase} {et~al.}(2015){Murase}, {Kashiyama}, {Kiuchi}, \&
  {Bartos}}]{2015ApJ...805...82M}
{Murase}, K., {Kashiyama}, K., {Kiuchi}, K., \& {Bartos}, I. 2015, \apj, 805,
  82

\bibitem[{{Murase} {et~al.}(2009){Murase}, {M{\'e}sz{\'a}ros}, \&
  {Zhang}}]{2009PhRvD..79j3001M}
{Murase}, K., {M{\'e}sz{\'a}ros}, P., \& {Zhang}, B. 2009, \prd, 79, 103001

\bibitem[{{Murase} {et~al.}(2018){Murase}, {Toomey}, {Fang}, {Oikonomou},
  {Kimura}, {Hotokezaka}, {Kashiyama}, {Ioka}, \&
  {M{\'e}sz{\'a}ros}}]{2018ApJ...854...60M}
{Murase}, K., {Toomey}, M.~W., {Fang}, K., {et~al.} 2018, \apj, 854, 60

\bibitem[{{Nicholl} {et~al.}(2017){Nicholl}, {Guillochon}, \&
  {Berger}}]{2017ApJ...850...55N}
{Nicholl}, M., {Guillochon}, J., \& {Berger}, E. 2017, \apj, 850, 55

\bibitem[{{Nicholl} {et~al.}(2018){Nicholl}, {Blanchard}, {Berger},
  {Alexander}, {Metzger}, {Bhirombhakdi}, {Chornock}, {Coppejans}, {Gomez},
  {Margalit}, {Margutti}, \& {Terreran}}]{Nicholl+18}
{Nicholl}, M., {Blanchard}, P.~K., {Berger}, E., {et~al.} 2018, \apjl, 866, L24

\bibitem[{{Olinto} {et~al.}(2017){Olinto}, {Adams}, {Aloisio}, {Anchordoqui},
  {Bergman}, {Bertaina}, {Bertone}, {Christl}, {Csorna}, {Eser}, {Fenu},
  {Hays}, {Hunter}, {Judd}, {Jun}, {Krizmanic}, {Kuznetsov}, {Martinez-Sierra},
  {Mastafa}, {Matthews}, {McEnery}, {Mitchell}, {Neronov}, {Otte}, {Parizot},
  {Paul}, {Perkins}, {Pr{\'e}v{\^o}t}, {Reardon}, {Reno}, {Sarazin},
  {Shinozaki}, {Stecker}, {Streitmatter}, {Wiencke}, \&
  {Young}}]{2017ICRC...35..542O}
{Olinto}, A.~V., {Adams}, J.~H., {Aloisio}, R., {et~al.} 2017, International
  Cosmic Ray Conference, 301, 542

\bibitem[{{Ostriker} \& {Gunn}(1969)}]{1969ApJ...157.1395O}
{Ostriker}, J.~P., \& {Gunn}, J.~E. 1969, \apj, 157, 1395

\bibitem[{{{\"O}zel} {et~al.}(2010){{\"O}zel}, {Psaltis}, {Ransom}, {Demorest},
  \& {Alford}}]{2010ApJ...724L.199O}
{{\"O}zel}, F., {Psaltis}, D., {Ransom}, S., {Demorest}, P., \& {Alford}, M.
  2010, \apjl, 724, L199

\bibitem[{{Perley} {et~al.}(2019){Perley}, {Mazzali}, {Yan}, {Cenko}, {Gezari},
  {Taggart}, {Blagorodnova}, {Fremling}, {Mockler}, {Singh}, {Tominaga},
  {Tanaka}, {Watson}, {Ahumada}, {Anupama}, {Ashall}, {Becerra}, {Bersier},
  {Bhalerao}, {Bloom}, {Butler}, {Copperwheat}, {Coughlin}, {De}, {Drake},
  {Duev}, {Frederick}, {Gonz{\'a}lez}, {Goobar}, {Heida}, {Ho}, {Horst},
  {Hung}, {Itoh}, {Jencson}, {Kasliwal}, {Kawai}, {Khanam}, {Kulkarni},
  {Kumar}, {Kumar}, {Kutyrev}, {Lee}, {Maeda}, {Mahabal}, {Murata}, {Neill},
  {Ngeow}, {Penprase}, {Pian}, {Quimby}, {Ramirez-Ruiz}, {Richer},
  {Rom{\'a}n-Z{\'u}{\~n}iga}, {Sahu}, {Srivastav}, {Socia}, {Sollerman},
  {Tachibana}, {Taddia}, {Tinyanont}, {Troja}, {Ward}, {Wee}, \&
  {Yu}}]{Perley+18}
{Perley}, D.~A., {Mazzali}, P.~A., {Yan}, L., {et~al.} 2019, \mnras, 484, 1031

\bibitem[{{Philippov} \& {Spitkovsky}(2018)}]{2018ApJ...855...94P}
{Philippov}, A.~A., \& {Spitkovsky}, A. 2018, \apj, 855, 94

\bibitem[{Pierog {et~al.}(2015)Pierog, Karpenko, Katzy, Yatsenko, \&
  Werner}]{PhysRevC.92.034906}
Pierog, T., Karpenko, I., Katzy, J.~M., Yatsenko, E., \& Werner, K. 2015, Phys.
  Rev. C, 92, 034906.
\newblock \url{https://link.aps.org/doi/10.1103/PhysRevC.92.034906}

\bibitem[{{Piro} \& {Kollmeier}(2016)}]{Piro&Kollmeier16}
{Piro}, A.~L., \& {Kollmeier}, J.~A. 2016, \apj, 826, 97

\bibitem[{{Planck Collaboration} {et~al.}(2016){Planck Collaboration}, {Ade},
  {Aghanim}, {Arnaud}, {Ashdown}, {Aumont}, {Baccigalupi}, {Banday},
  {Barreiro}, {Bartlett}, {Bartolo}, {Battaner}, {Battye}, {Benabed},
  {Beno{\^\i}t}, {Benoit-L{\'e}vy}, {Bernard}, {Bersanelli}, {Bielewicz},
  {Bock}, {Bonaldi}, {Bonavera}, {Bond}, {Borrill}, {Bouchet}, {Boulanger},
  {Bucher}, {Burigana}, {Butler}, {Calabrese}, {Cardoso}, {Catalano},
  {Challinor}, {Chamballu}, {Chary}, {Chiang}, {Chluba}, {Christensen},
  {Church}, {Clements}, {Colombi}, {Colombo}, {Combet}, {Coulais}, {Crill},
  {Curto}, {Cuttaia}, {Danese}, {Davies}, {Davis}, {de Bernardis}, {de Rosa},
  {de Zotti}, {Delabrouille}, {D{\'e}sert}, {Di Valentino}, {Dickinson},
  {Diego}, {Dolag}, {Dole}, {Donzelli}, {Dor{\'e}}, {Douspis}, {Ducout},
  {Dunkley}, {Dupac}, {Efstathiou}, {Elsner}, {En{\ss}lin}, {Eriksen},
  {Farhang}, {Fergusson}, {Finelli}, {Forni}, {Frailis}, {Fraisse},
  {Franceschi}, {Frejsel}, {Galeotta}, {Galli}, {Ganga}, {Gauthier}, {Gerbino},
  {Ghosh}, {Giard}, {Giraud-H{\'e}raud}, {Giusarma}, {Gjerl{\o}w},
  {Gonz{\'a}lez-Nuevo}, {G{\'o}rski}, {Gratton}, {Gregorio}, {Gruppuso},
  {Gudmundsson}, {Hamann}, {Hansen}, {Hanson}, {Harrison}, {Helou}, {Henrot-
  Versill{\'e}}, {Hern{\'a}ndez-Monteagudo}, {Herranz}, {Hildebrandt}, {Hivon},
  {Hobson}, {Holmes}, {Hornstrup}, {Hovest}, {Huang}, {Huffenberger}, {Hurier},
  {Jaffe}, {Jaffe}, {Jones}, {Juvela}, {Keih{\"a}nen}, {Keskitalo}, {Kisner},
  {Kneissl}, {Knoche}, {Knox}, {Kunz}, {Kurki-Suonio}, {Lagache},
  {L{\"a}hteenm{\"a}ki}, {Lamarre}, {Lasenby}, {Lattanzi}, {Lawrence}, {Leahy},
  {Leonardi}, {Lesgourgues}, {Levrier}, {Lewis}, {Liguori}, {Lilje},
  {Linden-V{\o}rnle}, {L{\'o}pez-Caniego}, {Lubin}, {Mac{\'\i}as-P{\'e}rez},
  {Maggio}, {Maino}, {Mandolesi}, {Mangilli}, {Marchini}, {Maris}, {Martin},
  {Martinelli}, {Mart{\'\i}nez-Gonz{\'a}lez}, {Masi}, {Matarrese}, {McGehee},
  {Meinhold}, {Melchiorri}, {Melin}, {Mendes}, {Mennella}, {Migliaccio},
  {Millea}, {Mitra}, {Miville-Desch{\^e}nes}, {Moneti}, {Montier}, {Morgante},
  {Mortlock}, {Moss}, {Munshi}, {Murphy}, {Naselsky}, {Nati}, {Natoli},
  {Netterfield}, {N{\o}rgaard-Nielsen}, {Noviello}, {Novikov}, {Novikov},
  {Oxborrow}, {Paci}, {Pagano}, {Pajot}, {Paladini}, {Paoletti}, {Partridge},
  {Pasian}, {Patanchon}, {Pearson}, {Perdereau}, {Perotto}, {Perrotta},
  {Pettorino}, {Piacentini}, {Piat}, {Pierpaoli}, {Pietrobon}, {Plaszczynski},
  {Pointecouteau}, {Polenta}, {Popa}, {Pratt}, {Pr{\'e}zeau}, {Prunet},
  {Puget}, {Rachen}, {Reach}, {Rebolo}, {Reinecke}, {Remazeilles}, {Renault},
  {Renzi}, {Ristorcelli}, {Rocha}, {Rosset}, {Rossetti}, {Roudier},
  {Rouill{\'e} d'Orfeuil}, {Rowan-Robinson}, {Rubi{\~n}o-Mart{\'\i}n},
  {Rusholme}, {Said}, {Salvatelli}, {Salvati}, {Sandri}, {Santos},
  {Savelainen}, {Savini}, {Scott}, {Seiffert}, {Serra}, {Shellard}, {Spencer},
  {Spinelli}, {Stolyarov}, {Stompor}, {Sudiwala}, {Sunyaev}, {Sutton},
  {Suur-Uski}, {Sygnet}, {Tauber}, {Terenzi}, {Toffolatti}, {Tomasi},
  {Tristram}, {Trombetti}, {Tucci}, {Tuovinen}, {T{\"u}rler}, {Umana},
  {Valenziano}, {Valiviita}, {Van Tent}, {Vielva}, {Villa}, {Wade}, {Wandelt},
  {Wehus}, {White}, {White}, {Wilkinson}, {Yvon}, {Zacchei}, \&
  {Zonca}}]{2016A&A...594A..13P}
{Planck Collaboration}, {Ade}, P.~A.~R., {Aghanim}, N., {et~al.} 2016, \aap,
  594, A13

\bibitem[{{Prentice} {et~al.}(2018){Prentice}, {Maguire}, {Smartt}, {Magee},
  {Schady}, {Sim}, {Chen}, {Clark}, {Colin}, {Fulton}, {McBrien}, {O'Neill},
  {Smith}, {Ashall}, {Chambers}, {Denneau}, {Flewelling}, {Heinze}, {Holoien},
  {Huber}, {Kochanek}, {Mazzali}, {Prieto}, {Rest}, {Shappee}, {Stalder},
  {Stanek}, {Stritzinger}, {Thompson}, \& {Tonry}}]{CowATLAS}
{Prentice}, S.~J., {Maguire}, K., {Smartt}, S.~J., {et~al.} 2018, \apj, 865, L3

\bibitem[{{Renault-Tinacci} {et~al.}(2018){Renault-Tinacci}, {Kotera},
  {Neronov}, \& {Ando}}]{Renault18}
{Renault-Tinacci}, N., {Kotera}, K., {Neronov}, A., \& {Ando}, S. 2018, \aap,
  611, A45

\bibitem[{{Rest} {et~al.}(2018){Rest}, {Garnavich}, {Khatami}, {Kasen},
  {Tucker}, {Shaya}, {Olling}, {Mushotzky}, {Zenteno}, {Margheim},
  {Strampelli}, {James}, {Smith}, {F{\"o}rster}, \& {Villar}}]{Rest+18}
{Rest}, A., {Garnavich}, P.~M., {Khatami}, D., {et~al.} 2018, Nature Astronomy,
  2, 307

\bibitem[{{Rivera Sandoval} {et~al.}(2018){Rivera Sandoval}, {Maccarone},
  {Corsi}, {Brown}, {Pooley}, \& {Wheeler}}]{RiveraSandoval+18}
{Rivera Sandoval}, L.~E., {Maccarone}, T.~J., {Corsi}, A., {et~al.} 2018,
  \mnras, 480, L146

\bibitem[{Shore {et~al.}(2009)Shore, Dall'Osso, \&
  Stella}]{10.1111/j.1365-2966.2008.14054.x}
Shore, S.~N., Dall'Osso, S., \& Stella, L. 2009, Monthly Notices of the Royal
  Astronomical Society, 398, 1869.
\newblock \url{https://doi.org/10.1111/j.1365-2966.2008.14054.x}

\bibitem[{{Smartt} {et~al.}(2018){Smartt}, {Clark}, {Smith}, {McBrien},
  {Maguire}, {O'Neil}, {Fulton}, {Magee}, {Prentice}, {Colin}, {Tonry},
  {Denneau}, {Stalder}, {Heinze}, {Weiland}, {Flewelling}, \&
  {Rest}}]{Smartt+18}
{Smartt}, S.~J., {Clark}, P., {Smith}, K.~W., {et~al.} 2018, The Astronomer's
  Telegram, 11727

\bibitem[{{Spitkovsky}(2006)}]{Spitkovsky06}
{Spitkovsky}, A. 2006, Astrophys. J. Lett., 648, L51

\bibitem[{{Stella} {et~al.}(2005){Stella}, {Dall'Osso}, {Israel}, \&
  {Vecchio}}]{2005ApJ...634L.165S}
{Stella}, L., {Dall'Osso}, S., {Israel}, G.~L., \& {Vecchio}, A. 2005, \apjl,
  634, L165

\bibitem[{Stella {et~al.}(2005)Stella, Dall{\textquotesingle}Osso, Israel, \&
  Vecchio}]{Stella_2005}
Stella, L., Dall{\textquotesingle}Osso, S., Israel, G.~L., \& Vecchio, A. 2005,
  The Astrophysical Journal, 634, L165.
\newblock \url{https://doi.org/10.1086%2F498685}

\bibitem[{{Svensson}(1987)}]{1987MNRAS.227..403S}
{Svensson}, R. 1987, \mnras, 227, 403

\bibitem[{{Tanaka} \& {Takahara}(2010)}]{2010ApJ...715.1248T}
{Tanaka}, S.~J., \& {Takahara}, F. 2010, \apj, 715, 1248

\bibitem[{{Tanaka} \& {Takahara}(2013)}]{2013MNRAS.429.2945T}
---. 2013, \mnras, 429, 2945

\bibitem[{{Tauris} {et~al.}(2015){Tauris}, {Langer}, \&
  {Podsiadlowski}}]{Tauris+15}
{Tauris}, T.~M., {Langer}, N., \& {Podsiadlowski}, P. 2015, \mnras, 451, 2123

\bibitem[{{The Pierre Auger Collaboration} {et~al.}(2017){The Pierre Auger
  Collaboration}, {Aab}, {Abreu}, {Aglietta}, {Albuquerque}, {Allekotte},
  {Almela}, {Alvarez Castillo}, {Alvarez- Mu{\~n}iz}, {Anastasi},
  {Anchordoqui}, {Andrada}, {Andringa}, {Aramo}, {Arsene}, {Asorey}, {Assis},
  {Aublin}, {Avila}, {Badescu}, {Balaceanu}, {Barbato}, {Barreira Luz},
  {Becker}, {Bellido}, {Berat}, {Bertaina}, {Bertou}, {Biermann}, {Biteau},
  {Blaess}, {Blanco}, {Blazek}, {Bleve}, {Boh{\'a}{\v{c}}ov{\'a}}, {Boncioli},
  {Bonifazi}, {Borodai}, {Botti}, {Brack}, {Brancus}, {Bretz}, {Bridgeman},
  {Briechle}, {Buchholz}, {Bueno}, {Buitink}, {Buscemi}, {Caballero-Mora},
  {Caccianiga}, {Caccianiga}, {Cancio}, {Canfora}, {Caramete}, {Caruso},
  {Castellina}, {Catalani}, {Cataldi}, {Cazon}, {Chavez}, {Chinellato},
  {Chudoba}, {Clay}, {Cobos}, {Colalillo}, {Coleman}, {Collica}, {Coluccia},
  {Concei{\c{c}}{\~a}o}, {Consolati}, {Consolati}, {Contreras}, {Cooper},
  {Coutu}, {Covault}, {Cronin}, {D'Amico}, {Daniel}, {Dasso}, {Daumiller},
  {Dawson}, {de Almeida}, {de Jong}, {De Mauro}, {de Mello Neto}, {De Mitri},
  {de Oliveira}, {de Souza}, {Debatin}, {Deligny}, {D{\'\i}az Castro}, {Diogo},
  {Dobrigkeit}, {D'Olivo}, {Dorosti}, {dos Anjos}, {Dova}, {Dundovic}, {Ebr},
  {Engel}, {Erdmann}, {Erfani}, {Escobar}, {Espadanal}, {Etchegoyen}, {Falcke},
  {Farmer}, {Farrar}, {Fauth}, {Fazzini}, {Fenu}, {Fick}, {Figueira},
  {Filip{\v{c}}i{\v{c}}}, {Freire}, {Fujii}, {Fuster}, {Ga{\"\i}or},
  {Garc{\'\i}a}, {Gat{\'e}}, {Gemmeke}, {Gherghel-Lascu}, {Ghia}, {Giaccari},
  {Giammarchi}, {Giller}, {G{\l}as}, {Glaser}, {Golup}, {G{\'o}mez Berisso},
  {G{\'o}mez Vitale}, {Gonz{\'a}lez}, {Gorgi}, {Grillo}, {Grubb}, {Guarino},
  {Guedes}, {Halliday}, {Hampel}, {Hansen}, {Harari}, {Harrison}, {Haungs},
  {Hebbeker}, {Heck}, {Heimann}, {Herve}, {Hill}, {Hojvat}, {Holt}, {Homola},
  {H{\"o}randel}, {Horvath}, {Hrabovsk{\'y}}, {Huege}, {Hulsman}, {Insolia},
  {Isar}, {Jandt}, {Johnsen}, {Josebachuili}, {Jurysek}, {K{\"a}{\"a}p{\"a}},
  {Kambeitz}, {Kampert}, {Keilhauer}, {Kemmerich}, {Kemp}, {Kemp},
  {Kieckhafer}, {Klages}, {Kleifges}, {Kleinfeller}, {Krause}, {Krohm},
  {Kuempel}, {Kukec Mezek}, {Kunka}, {Kuotb Awad}, {Lago}, {LaHurd}, {Lang},
  {Lauscher}, {Legumina}, {Leigui de Oliveira}, {Letessier-Selvon}, {Lhenry-
  Yvon}, {Link}, {Lo Presti}, {Lopes}, {L{\'o}pez}, {L{\'o}pez Casado},
  {Lorek}, {Luce}, {Lucero}, {Malacari}, {Mallamaci}, {Mandat}, {Mantsch},
  {Mariazzi}, {Mariș}, {Marsella}, {Martello}, {Martinez}, {Mart{\'\i}nez
  Bravo}, {Mas{\'\i}as Meza}, {Mathes}, {Mathys}, {Matthiae}, {Mayotte},
  {Mazur}, {Medina}, {Medina-Tanco}, {Melo}, {Menshikov}, {Merenda}, {Michal},
  {Micheletti}, {Middendorf}, {Miramonti}, {Mitrica}, {Mockler}, {Mollerach},
  {Montanet}, {Morello}, {Morlino}, {Mostaf{\'a}}, {M{\"u}ller}, {M{\"u}ller},
  {Muller}, {M{\"u}ller}, {Mussa}, {Naranjo}, {Nellen}, {Nguyen},
  {Niculescu-Oglinzanu}, {Niechciol}, {Niemietz}, {Niggemann}, {Nitz}, {Nosek},
  {Novotny}, {No{\v{z}}ka}, {N{\'u}{\~n}ez}, {Ochilo}, {Oikonomou}, {Olinto},
  {Palatka}, {Pallotta}, {Papenbreer}, {Parente}, {Parra}, {Paul}, {Pech},
  {Pedreira}, {P{\k{e}}kala}, {Pelayo}, {Pe{\~n}a-Rodriguez}, {Pereira},
  {Perlin}, {Perrone}, {Peters}, {Petrera}, {Phuntsok}, {Piegaia}, {Pierog},
  {Pimenta}, {Pirronello}, {Platino}, {Plum}, {Poh}, {Porowski}, {Prado},
  {Privitera}, {Prouza}, {Quel}, {Querchfeld}, {Quinn}, {Ramos-Pollan},
  {Rautenberg}, {Ravignani}, {Ridky}, {Riehn}, {Risse}, {Ristori}, {Rizi},
  {Rodrigues de Carvalho}, {Rodriguez Fernandez}, {Rodriguez Rojo},
  {Roncoroni}, {Roth}, {Roulet}, {Rovero}, {Ruehl}, {Saffi}, {Saftoiu},
  {Salamida}, {Salazar}, {Saleh}, {Salina}, {S{\'a}nchez}, {Sanchez-Lucas},
  {Santos}, {Santos}, {Sarazin}, {Sarmento}, {Sarmiento-Cano}, {Sato},
  {Schauer}, {Scherini}, {Schieler}, {Schimp}, {Schmidt}, {Scholten},
  {Schov{\'a}nek}, {Schr{\"o}der}, {Schr{\"o}der}, {Schulz}, {Schumacher},
  {Sciutto}, {Segreto}, {Shellard}, {Sigl}, {Silli}, {{\v{S}}m{\'\i}da},
  {Snow}, {Sommers}, {Sonntag}, {Soriano}, {Squartini}, {Stanca},
  {Stani{\v{c}}}, {Stasielak}, {Stassi}, {Stolpovskiy}, {Strafella}, {Streich},
  {Suarez}, {Suarez Dur{\'a}n}, {Sudholz}, {Suomij{\"a}rvi}, {Supanitsky},
  {{\v{S}}up{\'\i}k}, {Swain}, {Szadkowski}, {Taboada}, {Taborda}, {Theodoro},
  {Timmermans}, {Todero Peixoto}, {Tomankova}, {Tom{\'e}}, {Torralba Elipe},
  {Travnicek}, {Trini}, {Ulrich}, {Unger}, {Urban}, {Vald{\'e}s Galicia},
  {Vali{\~n}o}, {Valore}, {van Aar}, {van Bodegom}, {van den Berg}, {van
  Vliet}, {Varela}, {Vargas C{\'a}rdenas}, {V{\'a}zquez}, {Veberi{\v{c}}},
  {Ventura}, {Vergara Quispe}, {Verzi}, {Vicha}, {Villase{\~n}or}, {Vorobiov},
  {Wahlberg}, {Wainberg}, {Walz}, {Watson}, {Weber}, {Weindl}, {Wiede{\'n}ski},
  {Wiencke}, {Wilczy{\'n}ski}, {Winchen}, {Wirtz}, {Wittkowski}, {Wundheiler},
  {Yang}, {Yushkov}, {Zas}, {Zavrtanik}, {Zavrtanik}, {Zepeda}, {Zimmermann},
  {Ziolkowski}, {Zong}, \& {Zuccarello}}]{2017arXiv170806592T}
{The Pierre Auger Collaboration}, {Aab}, A., {Abreu}, P., {et~al.} 2017, ArXiv
  e-prints, arXiv:1708.06592

\bibitem[{Torres {et~al.}(2014)Torres, Cillis, Martín, \&
  de~Oña~Wilhelmi}]{TORRES201431}
Torres, D., Cillis, A., Martín, J., \& de~Oña~Wilhelmi, E. 2014, Journal of
  High Energy Astrophysics, 1-2, 31 .
\newblock
  \url{http://www.sciencedirect.com/science/article/pii/S2214404814000032}

\bibitem[{{Wang} {et~al.}(2007){Wang}, {Razzaque}, {M{\'e}sz{\'a}ros}, \&
  {Dai}}]{2007PhRvD..76h3009W}
{Wang}, X.-Y., {Razzaque}, S., {M{\'e}sz{\'a}ros}, P., \& {Dai}, Z.-G. 2007,
  \prd, 76, 083009

\bibitem[{{Woosley}(2010)}]{Woosley10}
{Woosley}, S.~E. 2010, Astrophys. J. Lett., 719, L204

\bibitem[{{Yu} {et~al.}(2013){Yu}, {Zhang}, \& {Gao}}]{Yu+13}
{Yu}, Y.-W., {Zhang}, B., \& {Gao}, H. 2013, Astrophys. J. Lett., 776, L40

\bibitem[{{Zhang} {et~al.}(2018){Zhang}, {Murase}, {Kimura}, {Horiuchi}, \&
  {M{\'e}sz{\'a}ros}}]{2018PhRvD..97h3010Z}
{Zhang}, B.~T., {Murase}, K., {Kimura}, S.~S., {Horiuchi}, S., \&
  {M{\'e}sz{\'a}ros}, P. 2018, \prd, 97, 083010

\end{thebibliography}

\end{document}